\documentclass{aa}  
\def \eg {e.g.}

\def \ie {i.e.}
\def \cf {cf.}
\def \lcdm {{\hbox{$\Lambda$CDM}}}
\def \omegam {{\hbox{$\Omega_m$}}}
\def \omegal {{\hbox{$\Omega_\Lambda$}}}
\def \hzero {{\hbox{$H_0$}}}
\def \arcmin {\hbox{$^\prime$}}
\def \arcsec {\hbox{$^{\prime\prime}$}}
\def \deg {\hbox{$^\circ$}}
\def \nh {\hbox{$N_{\rm H}$}}

\def \ktd {{\hbox{$kT_d$}}}
\def \mach {{\hbox{$\mathcal{M}$}}}

\def \compr {{\hbox{$\mathcal{C}$}}}

\def \zsun {\hbox{${\rm Z_\odot}$}}

\def \mfive {\hbox{$M_{500}$}}

\newcommand{\kmsmpc }{\mbox{km s$^{-1}$ Mpc$^{-1}$}}

\newcommand{\kev }{\mbox{keV}}

\newcommand{\jy }{\mbox{Jy}}

\newcommand{\muG }{\mbox{$\mu$G}}

\newcommand{\obsid }{ObsID}

\newcommand{\ciao }{\textsc{ciao}}

\newcommand{\caldb }{\textsc{caldb}}

\newcommand{\esas }{\textsc{esas}}

\newcommand{\sas }{\textsc{sas}}

\newcommand{\proffit }{\textsc{proffit}}

\newcommand{\xmm }{{\em XMM-Newton}}
\newcommand{\chandra }{{\em Chandra}}

\newcommand{\fermi }{{\em Fermi}}

\newcommand{\suzaku }{{\em Suzaku}}

\usepackage{graphicx}
\usepackage{amssymb}
\usepackage{multirow,bigdelim}
\usepackage{txfonts}
\usepackage[export]{adjustbox}
\usepackage{color}

\begin{document} 

\title{Shock acceleration efficiency in radio relics}

\authorrunning{A. Botteon, G. Brunetti, D. Ryu, S. Roh} 
\titlerunning{Shock acceleration efficiency in radio relics}

\author{A. Botteon\inst{1,2,3}, G. Brunetti\inst{2}, D. Ryu\inst{4} and S. Roh\inst{4}}

\institute{
Dipartimento di Fisica e Astronomia, Universit\`{a} di Bologna, via P.~Gobetti 93/2, I-40129 Bologna, Italy \\
\email{botteon@ira.inaf.it}
\and
INAF - IRA, via P.~Gobetti 101, I-40129 Bologna, Italy 
\and
Leiden Observatory, Leiden University, PO Box 9513, NL-2300 RA Leiden, The Netherlands
\and
Department of Physics, School of Natural Sciences, UNIST, Ulsan 44919, Korea
}

\date{Received XXX; accepted YYY}

\abstract
{Radio relics in galaxy clusters are giant diffuse synchrotron sources powered in cluster outskirts by merger shocks. Although the relic--shock connection has been consolidated in recent years by a number of observations, the details of the mechanisms leading to the formation of relativistic particles in this environment are still not well understood.}
{The diffusive shock acceleration (DSA) theory is a commonly adopted scenario to explain the origin of cosmic rays at astrophysical shocks, including those in radio relics in galaxy clusters. However, in a few specific cases it has been shown that the energy dissipated by cluster shocks is not enough to reproduce the luminosity of the relics via DSA of thermal particles. Studies based on samples of radio relics are required to further address this limitation of the mechanism.}
{In this paper, we focus on ten well-studied radio relics with underlying shocks observed in the X-rays and calculate the electron acceleration efficiency of these shocks that is necessary to reproduce the observed radio luminosity of the relics.}
{We find that in general the standard DSA cannot explain the origin of the relics if electrons are accelerated from the thermal pool with an efficiency significantly smaller than 10\%. Our results show that other mechanisms, such as shock re-acceleration of supra-thermal seed electrons or a modification of standard DSA, are required to explain the formation of radio relics.}
{}


\maketitle
%

\section{Introduction}

Astrophysical shock waves are able to accelerate particles over a broad range of scales, from astronomical units in the Sun heliosphere to megaparsecs in clusters of galaxies. Among the numerous physical processes proposed, the diffusive shock acceleration (DSA) theory provides a general explanation of particle acceleration in several astrophysical environments \citep[\eg][]{blandford87rev}. This process is based on the original idea of \citet{fermi49}, according to which particles are scattered upstream and downstream of the shock by plasma irregularities, gaining energy at each reflection. \\
\indent
Radio relics in galaxy clusters are giant synchrotron sources that are explained assuming that relativistic particles are accelerated by shocks crossing the intra-cluster medium (ICM) \citep[\eg][]{ensslin98relics, roettiger99a3667}. Whilst the relic--shock connection is nowadays well consolidated by radio and X-ray observations \citep[see][for reviews]{brunetti14rev, vanweeren19rev}, the details of the acceleration mechanisms are still not fully understood. \\
\indent
To date, the acceleration efficiency of cosmic rays (CR) at astrophysical shocks is mainly constrained by studies of supernova remnants (SNR) in our Galaxy, where strong shocks ($\mach \sim 10^3$) propagate in a low-beta plasma ($\beta_{pl} = P_{th}/P_{B}$, \ie,\ the ratio between the thermal and magnetic pressures) medium and are able to transfer $\sim10\%$ or more of the energy flux through them into cosmic ray protons (CRp), and a smaller fraction into cosmic ray electrons \citep[CRe; \eg,][]{jones11, morlino12, caprioli14a, caprioli15, park15}. In contrast, radio relics in the outskirts of merging galaxy clusters probe particle acceleration in action at much weaker shocks ($\mach \lesssim 3-5$) and in a high-$\beta_{pl}$ environment such as the ICM, where the thermal pressure dominates over the magnetic pressure. In this case, the acceleration efficiency of CRp is still poorly understood, although current models predict efficiencies that are less than a few percent \citep[\eg,][]{kang05, kang13}, in agreement with direct constraints coming from $\gamma$-ray observations of galaxy clusters \citep[\eg,][]{ackermann10, ackermann14, ackermann16}. On the other hand, the observed connection between radio relics and shocks in merging galaxy clusters demonstrates that electron acceleration (or re-acceleration) at these shocks is efficient, implying surprisingly large values of the ratio of CRe/CRp acceleration efficiencies if these particles are extracted from shocks by the same population (\eg,\ the thermal ICM). This poses fundamental questions on the mechanisms leading to the formation of relativistic particles in radio relics \citep[\eg,][]{vazza14challenge, vazza15efficiency, vazza16}. \\
\indent
In recent years, deep X-ray observations performed with \chandra, \xmm, and \suzaku\ have led to an increase in the number of shocks detected in merging galaxy clusters \citep[\eg,][for recent works]{akamatsu17a2255, canning17, emery17, botteon18edges, tholken18, urdampilleta18}. In a few cases, when the shock front is co-spatially located with a radio relic, it has been shown (under reasonable assumptions on the minimum momentum of the accelerated electrons) that DSA is severely challenged by the large acceleration efficiencies required to reproduce the total radio luminosity of the relics if particles are accelerated from the ICM thermal pool \citep{botteon16a115, eckert16a2744, hoang17}. \\
To mitigate the problem of the high acceleration efficiencies implied by cluster shocks, recent theoretical models assume a pre-existing population of CRe at the position of the relic that is re-accelerated by the passage of the shock \citep[\eg,][]{markevitch05, macario11, kang11, kang12relics, kang14, pinzke13}. This re-acceleration scenario seems to be supported by the observation of radio galaxies located nearby or within a number of radio relics \citep[\eg,][]{bonafede14reacc, shimwell15, botteon16a115, vanweeren17a3411, digennaro18sausage}. \\
\indent
In order to test the scenario of shock acceleration of thermal particles as the origin of radio relics, in this paper, for the first time we evaluate the efficiency of particle acceleration at cluster shocks using a homogeneous approach in a relatively large number of radio relics. We adopt a \lcdm\ cosmology with $\omegal = 0.7$, $\omegam = 0.3$ and $\hzero = 70$ \kmsmpc\ throughout.

\section{Theoretical framework}

\subsection{Computation of the acceleration efficiency in radio relics}\label{sec:computation}

The electron acceleration efficiency $\eta_e$ can be related to the observed synchrotron luminosity that is produced by the shock-accelerated electrons. In particular, $\eta_e$ is evaluated assuming that a fraction of the kinetic energy flux from a shock with speed $V_{sh}$, surface $A$, compression factor \compr, and upstream mass density $\rho_u$ is converted into CRe acceleration to produce the bolometric ($\geq \nu_0$) synchrotron luminosity of the relic

\begin{equation}\label{eq:luminosity}
\int_{\nu_0} L(\nu)\,{\rm d}\nu \simeq \frac{1}{2} A \rho_u V_{sh}^3 \eta_e \Psi \left( 1 - \frac{1}{\compr^2} \right) {\frac{ B^2 }{B_{cmb}^2 + B^2}}
,\end{equation}

\noindent
where $\frac{ B^2 }{B_{cmb}^2 + B^2}$ takes into account that the energy of electrons is radiated away via both synchrotron and inverse Compton emission ($B$ is the intensity of the magnetic field and $B_{cmb} = 3.25(1+z)^2$ \muG\ is the equivalent magnetic field of the cosmic microwave background at redshift $z$), the term

\begin{equation}\label{eq:psi}
\Psi = {\frac{\int_{p_{0}} N_{inj}(p) E\,{\rm d}p}{\int_{p_{min}} N_{inj}(p) E\,{\rm d}p}}
\end{equation}

\noindent
accounts for the ratio of the energy flux  injected in ``all'' electrons and those visible in the radio band ($\nu \geq \nu_0$), $N_{inj}(p)$ is the electron momentum distribution, $p_0 \simeq 10^3 m_e c \sqrt{\frac{\nu_0 {[\rm MHz]}/(1+z)}{4.6 B[\rm \mu G]}}$ is the momentum of the relativistic electrons emitting the synchrotron frequency $\nu_0$, and $p_{min}$ is the minimum momentum of accelerated electrons.

\begin{figure}[t]
 \centering
 \includegraphics[width=\hsize,trim={1.6cm 6.6cm 2.6cm 4.3cm},clip]{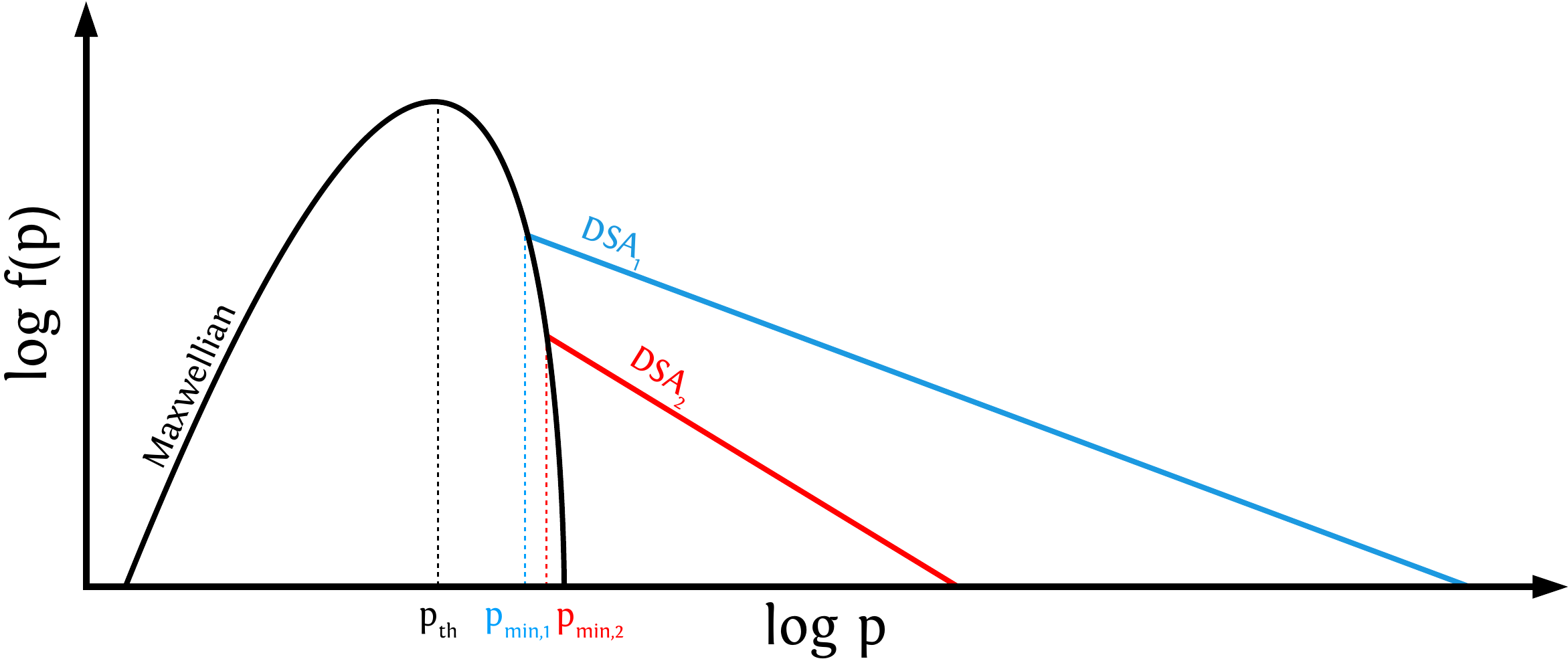}
 \caption{Schematic representation of the electron momentum distribution in a downstream region. The two power laws show the DSA spectra (Eq.~\ref{eq:delta-inj}) in the case of two Mach numbers $\mach_1$ (\textit{blue}) $>$ $\mach_2$ (\textit{red}).}
 \label{fig:definition}
\end{figure}

We assume that a fraction $\eta_e$ of the energy flux that can be dissipated at the shock surface\footnote{That is the change in the kinetic energy flux across the shock; \eg,\ \citet{finoguenov10}.} is channelled into the accelerated relativistic electrons in the downstream region $F_{relic}$:

\begin{equation}\label{eq:conservation}
 \overbrace{\frac{1}{2} V_{sh}^{3} \rho_u \left( 1 - \frac{1}{\compr^2} \right)}^{\Delta F_{KE}} \eta_e  = \overbrace{V_d \epsilon_{e,d}}^{F_{relic}}
,\end{equation}

\noindent
where

\begin{equation}\label{eq:v_down}
 V_d = c_{s,u} \frac{\mach^2 +3}{4 \mach}
,\end{equation}

\noindent
which is the downstream velocity, $c_{s,u}$ is the upstream sound speed, and

\begin{equation}\label{eq.pdflion_e}
 \epsilon_{e,d} = \int_{p_{min}} N_{inj}(p) E\,{\rm d}p
,\end{equation}

\noindent
which is the downstream energy density of the accelerated electrons. Although the acceleration efficiency can be computed for any possible electron momentum distribution, in the following we adopt a power-law momentum distribution of CRe in the form

\begin{equation}\label{eq:powerlaw}
 N_{inj}(p) = K_e p^{-\delta_{inj}}
,\end{equation}

\noindent
which is motivated by the single power laws generally used to describe the integrated radio spectra of radio relics \citep[\eg,][]{vanweeren19rev} and by DSA (see following section). If electrons are accelerated from the thermal pool starting from a minimum momentum as shown in Fig.~\ref{fig:definition}, a relationship between the minimum momentum and the normalization of the spectrum can be derived by matching the number density of nonthermal electrons with that of thermal electrons with momentum $p_{min}$ assuming $K_e p_{min}^{-\delta_{inj}}$. This leads to

\begin{equation}\label{eq:k_e}
 K_e = \frac{4}{\sqrt{\pi}} n_d \frac{p_{min}^{2+\delta_{inj}}}{p_{th}^3} \exp \left[ - \left( \frac{p_{min}}{p_{th}} \right)^2 \right]
,\end{equation}

\noindent
where $p_{th}=\sqrt{2 m_e \ktd}$ is the electron thermal peak momentum in the downstream gas (Fig.~\ref{fig:definition}). At this point, it is important to know what kind of acceleration efficiency (or parameters $p_{min}$, $K_e$) is necessary to generate the observed radio properties of radio relics. To address this question we combine Eqs.~\ref{eq:luminosity}, \ref{eq:conservation}, and \ref{eq:k_e}, and obtain

\begin{equation}\label{eq:p_min}
 p_{min}^{2+\delta_{inj}} \exp \left[ - \left( \frac{p_{min}}{p_{th}} \right)^2 \right] = \frac{\sqrt{\pi}}{4} \frac{\int_{\nu_0} L(\nu)\,{\rm d}\nu}{n_d V_d A} \frac{p_{th}^3}{ \int_{p_0} p^{-\delta_{inj}} E \,{\rm d}p} \frac{B_{cmb}^2 + B^2}{B^2}
,\end{equation}

\noindent
which can be used to determine the minimum momentum of the electrons (or the efficiency) that is necessary to generate an observed radio emission for a given set of shock and relic parameters (namely Mach number, density, temperature, surface area, and magnetic field). The surface of the shock is assumed to be $A=\pi R^2$, where $R$ is the semi-axis of the relic emission crossed by the shock.

\begin{figure}[t]
 \centering
 \includegraphics[width=.48\textwidth]{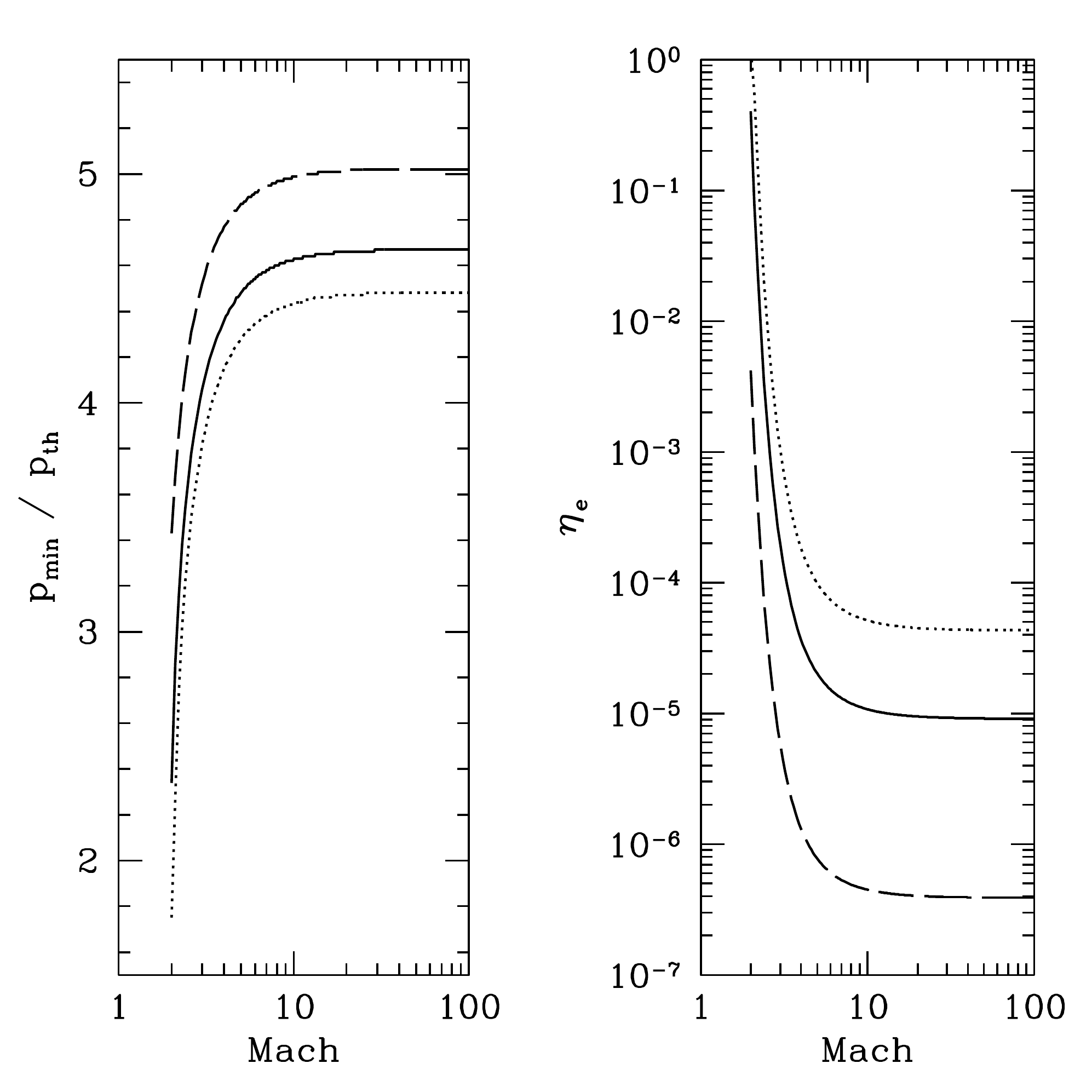}
 \includegraphics[width=.48\textwidth]{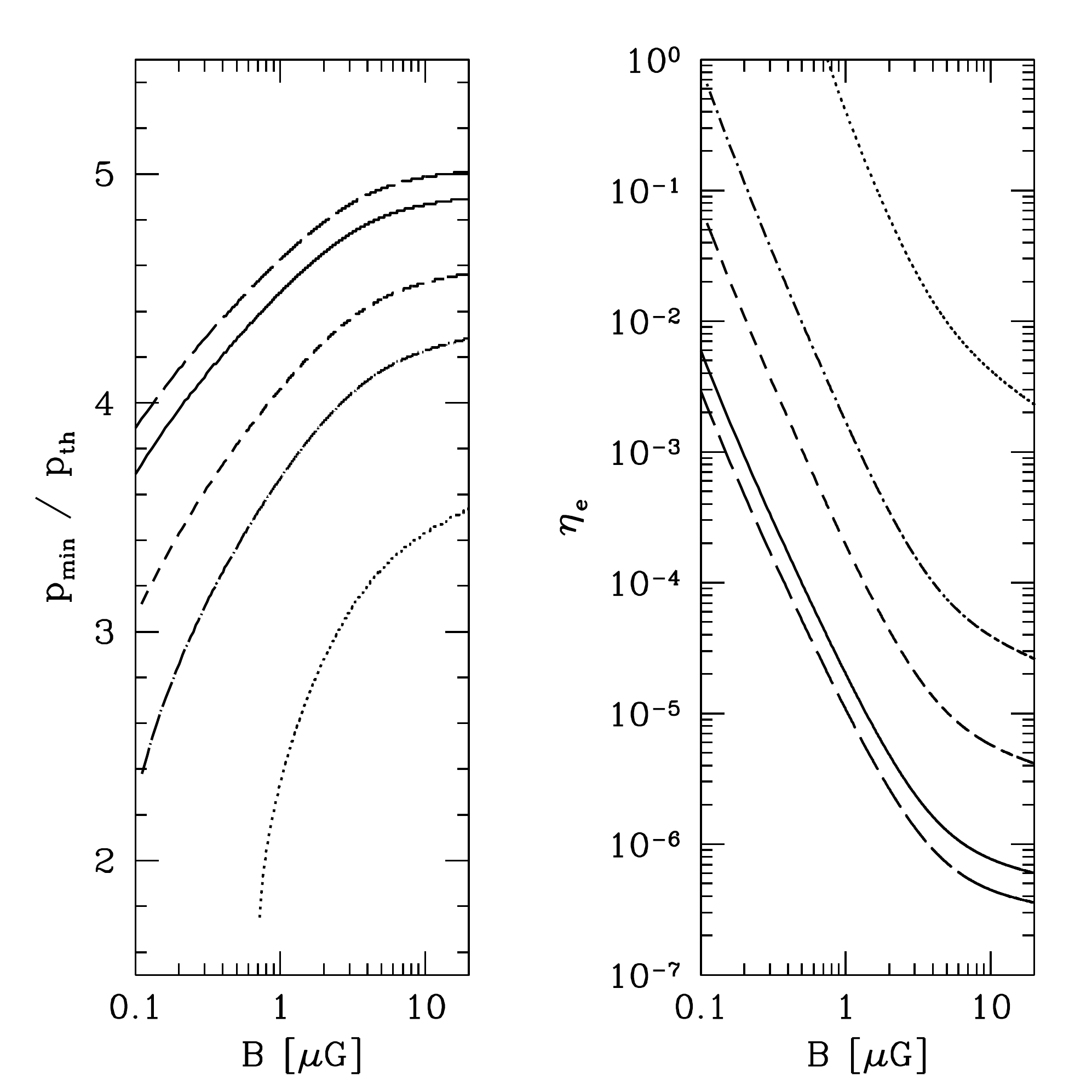}
 \caption{Values of $p_{min}/p_{th}$ and $\eta_e$ for a mock radio relic (see text) at fixed magnetic field (\textit{top}) and Mach number (\textit{bottom}). In the top panels, curves denote different values of $B$: 0.5 \muG\ (\textit{dotted}), 1 \muG\ (\textit{solid}) and 10 \muG\ (\textit{dashed}). In the bottom panels, curves denote different values of \mach: 2 (\textit{dotted}), 2.5 (\textit{dot-dashed}), 3 (\textit{short dashed}), 5 (\textit{solid}) and 10 (\textit{long dashed}).}
 \label{fig:efficiency_mock}
\end{figure}

Our knowledge of $B$ in clusters is poor and only a few constraints on the field strength in relics are available in the literature. In particular, the magnetic fields can be boosted at some level by shock compression and/or amplification in these dynamically active regions \citep{bonafede13, ji16}, perhaps reaching values up to 5 \muG\ \citep[\eg,][]{vanweeren10, botteon16gordo, rajpurohit18}. This is important to keep in mind because the required acceleration efficiency estimated with our approach is smaller for higher magnetic fields (see below). 

\subsection{Basic DSA relations}

Among the possible mechanisms that can produce a power-law distribution of CRe, DSA is customarily used to describe particle acceleration at astrophysical shocks and radio relics. In this scenario, the slope $\delta_{inj}$ of the accelerated electrons (\ie,\ the injection spectrum) in Eq.~\ref{eq:powerlaw} is 

\begin{equation}\label{eq:delta-inj}
 \delta_{inj} = 2\frac{\mach^2 + 1}{\mach^2-1}
,\end{equation}

\noindent
and it depends only on the shock Mach number \citep[e.g.,][]{blandford87rev}. \\
\indent
Under stationary conditions and assuming that the physical conditions in the downstream regions do not change with distance from the shock, the electron spectrum integrated in the downstream region follows a power law with a slope of $\delta = \delta_{inj} + 1.$ Thus, the integrated synchrotron spectrum is connected with the shock Mach number via

\begin{equation}\label{eq:alpha-mach}
 \alpha =  \frac{\mach^2 + 1}{\mach^2 -1} \equiv \alpha_{inj} + \frac{1}{2}\:.
\end{equation}

\noindent
As a consequence of the above relations, DSA predicts that for strong shocks ($\mach \rightarrow \infty$) the asymptotic behavior of the spectral index is $\alpha \rightarrow 1$ (and $\alpha_{inj} \rightarrow 0.5$), while for weak shocks ($\mach \lesssim 3-5$) it is $\alpha > 1$ (and $\alpha_{inj} > 0.5$). 

\begin{table*}[h]
 \centering
 \caption{Sample of galaxy clusters with radio relics and detected underlying shocks. Reported values of \mfive\ (mass within a radius that encloses a mean overdensity of 500 with respect to the critical density at the cluster redshift) and \nh\ (hydrogen column density) are taken from \citet{planck14xxix} and \citet{willingale13}, respectively. For the Sausage Cluster, \mfive\ is from \citet{degasperin14}. Redshifts are taken from the NASA/IPAC Extragalactic Database (NED).}
 \label{tab:sample_eff}
  \begin{tabular}{lccccccccc} 
  \hline
  Cluster name & RA$_{\rm{J}2000}$ & DEC$_{\rm{J}2000}$ & \mfive\ & $z$ & \nh\ & Instrument & \multicolumn{2}{c}{Ref.} \\
  & (h,m,s) & ($\deg$,\arcmin,\arcsec) & ($10^{14}$ M$_\odot$) & & ($10^{20}$ cm$^{-2}$) & & R & X \\
  \hline
  A2744 & 00 14 19 & $-$30 23 22 & 9.56 & 0.308 & 1.46 & \xmm\ & $^1$ & $^2$ \\
  A115 & 00 55 60 & $+$26 22 41 & 7.20 & 0.197 & 6.38 & \chandra & $^3$ & $^3$  \\
  El Gordo & 01 02 53 & $-$49 15 19 & 8.80 & 0.870 & 1.78 & \chandra &  $^4$ &  $^4$ \\
  A521 & 04 54 09 & $-$10 14 19 & 6.90 & 0.253 & 6.06 & \chandra &  $^5$ & $^6$ \\
  A3376 & 06 01 45 & $-$39 59 34 & 2.27 & 0.046 & 5.81 & \xmm &  $^7$ & $^8$ \\
  Toothbrush Cluster & 06 03 13 & $+$42 12 31 & 11.1 & 0.225 & 33.4 & \chandra & $^9$ & $^{10}$ \\
  Bullet Cluster & 06 58 31 & $-$55 56 49 & 12.4 & 0.296 & 6.43 & \chandra & $^{11}$ & $^{12}$ \\
  RXC J1314.4-2515 & 13 14 28 & $-$25 15 41 & 6.15 & 0.247 & 8.91 & \xmm & $^{13}$ & $^{14}$ \\
  A2146 & 15 56 09 & $+$66 21 21 & 3.85 & 0.234 & 3.35 & \chandra & $^{15}$ & $^{16}$ \\
  Sausage Cluster & 22 42 53 & $+$53 01 05 & 7.97 & 0.192 & 46.4 & \chandra & $^{17}$ & $^{18}$ \\
  \hline
  \multicolumn{9}{{p{.85\textwidth}}}{\textit{Notes.} References (R=radio, X=X-ray): $^1$Giacintucci et al.~(in preparation); $^2$\citet{eckert16a2744}; $^3$\citet{botteon16a115}; $^4$\citet{botteon16gordo}; $^5$\citet{giacintucci08}; $^6$\citet{bourdin13}; $^7$\citet{kale12a3376}; $^8$\citet{urdampilleta18}; $^9$\citet{vanweeren12toothbrush}; $^{10}$\citet{vanweeren16toothbrush}; $^{11}$\citet{shimwell14}; $^{12}$\citet{shimwell15}; $^{13}$\citet{venturi13}; $^{14}$\citet{mazzotta11}; $^{15}$\citet{russell12}; $^{16}$\citet{hoang19a2146}; $^{17}$\citet{vanweeren10}; $^{18}$\citet{akamatsu15}.}
  \end{tabular}
\end{table*}

\begin{table*}[h]
 \centering
 \caption{Parameters used to compute the acceleration efficiency. Values of the downstream temperature \ktd\ and radio flux density $S$ are derived from the works listed in Table~\ref{tab:sample_eff}.}
 \label{tab:efficiency_param}
  \begin{tabular}{lccccccc} 
  \hline
  Cluster name & Position & \compr\ & $n_d$ & \ktd\ & $A$ & $S_\nu$ & $\nu$ \\
  & & & (cm$^{-3}$) & (\kev) & ($\pi$ kpc$^2$) & (m\jy) & (MHz) \\
  \hline
  A2744$^{\dagger}$ & NE & $1.90^{+0.60}_{-0.40}$  & $1.8\times10^{-4}$ & 12.3 & $740^2$ & 20 & 1400 \\
  A115 & N & $2.15^{+0.16}_{-0.14}$ & $1.5\times10^{-3}$ & 7.9 & $180^2$ & 34 & 1400 \\
  El Gordo & NW & $2.88^{+0.30}_{-0.25}$ & $8.5\times10^{-4}$ & 17.9 & $350^2$ & 28 & 610 \\
  A521 & SE & $2.41^{+0.71}_{-1.41}$ & $3.0\times10^{-4}$ & 7.0 & $490^2$ & 42 & 610 \\
  A3376 & E & $1.98^{+0.27}_{-0.30}$ & $9.0\times10^{-4}$ & 4.7 & $260^2$ & 40 & 1400 \\
  Toothbrush Cluster & N & $1.37^{+0.18}_{-0.17}$ & $5.5\times10^{-4}$ & 8.2 & $300^2$ & 480 & 610 \\
  Bullet Cluster & E & $2.15^{+0.16}_{-0.14}$ & $5.0\times10^{-4}$ & 13.5 & $250^2$ & 5 & 2100 \\
  RXC J1314.4-2515 & W & $1.96^{+0.42}_{-0.36}$ & $1.0\times10^{-3}$ & 13.5 & $330^2$ & 85 & 325 \\
  A2146 & NW & $1.69^{+0.06}_{-0.06}$ & $3.5\times10^{-3}$ & 14.5 & $160^2$ & 0.8 & 1500 \\
  Sausage Cluster & N & n.a. & $3.0\times10^{-4}$ & 8.5 & $900^2$ & 337 & 610  \\
  \hline
  \multicolumn{8}{{p{0.7\textwidth}}}{\textit{Notes.} $^{\dagger}$Compression factor and downstream density taken from \citet{eckert16a2744}.}
  \end{tabular}
\end{table*} 

Following the \textit{thermal leakage injection} scenario for CRp \citep[e.g.,][]{gieseler00}, we assume here that only electrons with a minimum momentum threshold, $p_{min}$, can be accelerated. For electron acceleration at weak shocks, the physical details that determine $p_{min}$ are still poorly known \citep[e.g.,][]{guo14a, kang19}; consequently, we use $p_{min}$ as a free parameter that is connected with the efficiency (a larger $p_{min}$ corresponds to lower efficiency). Following the framework described in Section~\ref{sec:computation} and the DSA relations, in Fig.~\ref{fig:efficiency_mock} we report an example where we plot the $p_{min}/p_{th}$ and $\eta_e$ for a hypothetical radio relic at $z=0.1$ with a favorable combination of $\ktd = 10$ \kev, $n_d = 1.0 \times 10^{-3} $ cm$^{-3}$, $S_{1.4\:\rm{GHz}} = 5$ m\jy, and $A=750^2 \pi$ kpc$^2$ for different values of the Mach number and the magnetic field strength. Figure~\ref{fig:efficiency_mock} immediately identifies the problem: despite the optimistic parameters, these plots already demonstrate that DSA of thermal electrons becomes problematic (\ie,\ high $\eta_e$ or large $B$ are required) for weak shocks, that is, those of typically $\mach \lesssim 2-2.5$, and shocks of this strength are relatively common in the ICM and in radio relics. This is because, for weak shocks, an increasingly large fraction of the energy of the accelerated electrons is dumped into subGeV particles (formally, for $\mach < 2.2$, the majority of the energy is piled up into subrelativistic electrons).

\section{Sample of radio relics with underlying shocks}\label{sec:sample}

We select a sample of ten radio relics with underlying shocks observed in the X-rays. The clusters are listed in Table~\ref{tab:sample_eff} and include a few double radio relic systems. The sample is composed of well-studied radio relics with good radio and X-ray data available, which are essential to determine the spectral index of the relics and the properties of the underlying shocks. In particular, the detection of a shock co-spatially coincident with the relic (or at least a part of it) is necessary to evaluate the particle acceleration efficiency. \\
We point out that the well-known double radio relic system in A3667 \citep{rottgering97, johnstonhollitt03} is not considered here because the measured spectral indexes of the two radio relics are $\leq 1$ \citep{hindson14, riseley15}, and are therefore already in tension with DSA from the thermal pool (which would approach $\alpha = 1$ for very strong shocks; see Eq.~\ref{eq:alpha-mach}). \\
\indent
We derive the relevant quantities reported in Table~\ref{tab:efficiency_param} that are necessary to compute the acceleration efficiency of electrons for all the shocks associated with the radio relics in our sample. In particular, we start from downstream quantities (\ie,\ temperature and density), usually better constrained by X-ray observations, and from the Mach number of the shock derived from the density jump to derive upstream quantities using the Rankine-Hugoniot jump conditions \citep{landau59}. Downstream temperatures are taken from the literature while downstream densities and compression factors are obtained from the re-analysis of the surface brightness profiles extracted across the shocks. Our broken power-law fit confirms previous analyses of the same targets, providing evidence of a discontinuity coincident with the outer edge of radio emission for nine out of ten relics in the sample. The only case where we use a single power-law model to fit the surface brightness profile is for the Sausage relic, which is known to not exhibit a surface brightness jump across its surface. Details of the analysis including X-ray images and sectors used to extract the surface brightness profiles are given in Appendix~\ref{app:param}.

\begin{table*}[h]
 \centering
 \caption{Observed X-ray Mach number derived from the surface brightness analysis ($\mach_X$) and integrated spectral index from the literature ($\alpha_{radio}$). These were used to compute the expected integrated spectral index ($\alpha_{DSA}$) and Mach number ($\mach_{DSA}$) from DSA (Eq.~\ref{eq:alpha-mach}). References for the integrated spectral indexes are also listed.}
 \label{tab:x_vs_radio}
  \begin{tabular}{lcccccc} 
  \hline
  Cluster name & Position & $\mach_X$ & $\mach_{DSA}$ & $\alpha_{radio}$ & $\alpha_{DSA}$ & Reference \\
  \hline
  A2744 & NE & $1.65^{+0.59}_{-0.31}$ & $2.69^{+0.42}_{-0.27}$ & $1.32^{+0.09}_{-0.09}$ & $2.61^{+1.35}_{-0.66}$ & \citet{pearce17} \\
  A115 & N & $1.87^{+0.16}_{-0.13}$ & $4.58^{+\infty}_{-2.50}$ & $1.10^{+0.50}_{-0.50}$ & $1.80^{+0.19}_{-0.16}$ & \citet{govoni01six} \\
  El Gordo & NW & $2.78^{+0.63}_{-0.38}$ & $2.53^{+1.04}_{-0.41}$ & $1.37^{+0.20}_{-0.20}$ & $1.30^{+0.12}_{-0.11}$ & \citet{botteon16gordo} \\
  A521 & SE & $2.13^{+1.13}_{-1.13}$ & $2.33^{+0.05}_{-0.04}$ & $1.45^{+0.02}_{-0.02}$ & $1.57^{+\infty}_{-0.36}$ & \citet{macario13} \\
  A3376 & E & $1.71^{+0.25}_{-0.24}$ &  $2.53^{+0.28}_{-0.20}$ & $1.37^{+0.08}_{-0.08}$ & $2.04^{+0.68}_{-0.34}$ & \citet{george15} \\
  Toothbrush Cluster & N & $1.25^{+0.13}_{-0.12}$ & $3.79^{+0.26}_{-0.22}$ & $1.15^{+0.02}_{-0.02}$ & $4.56^{+3.67}_{-1.34}$ & \citet{rajpurohit18} \\
  Bullet Cluster & E & $1.87^{+0.16}_{-0.13}$ & $2.01^{+0.19}_{-0.14}$ & $1.66^{+0.14}_{-0.14}$ & $1.80^{+0.19}_{-0.16}$ & \citet{shimwell15} \\
  RXC J1314.4-2515 & W & $1.70^{+0.40}_{-0.28}$ & $3.18^{+0.87}_{-0.45}$ & $1.22^{+0.09}_{-0.09}$ & $2.06^{+0.91}_{-0.47}$ & \citet{george17}\\
  A2146 & NW & $1.48^{+0.05}_{-0.05}$ & $3.91^{+1.95}_{-0.73}$ & $1.14^{+0.08}_{-0.08}$ & $2.68^{+0.23}_{-0.19}$ & \citet{hoang19a2146} \\
  Sausage Cluster & N & n.a. & $4.38^{+1.06}_{-0.59}$ & $1.11^{+0.04}_{-0.04}$ & n.a. & \citet{hoang17} \\
  \hline
  \end{tabular}
\end{table*}

\begin{figure*}[h!]
 \centering
 \includegraphics[width=\textwidth,trim={1.3cm 4.3cm 1.5cm 4.65cm},clip]{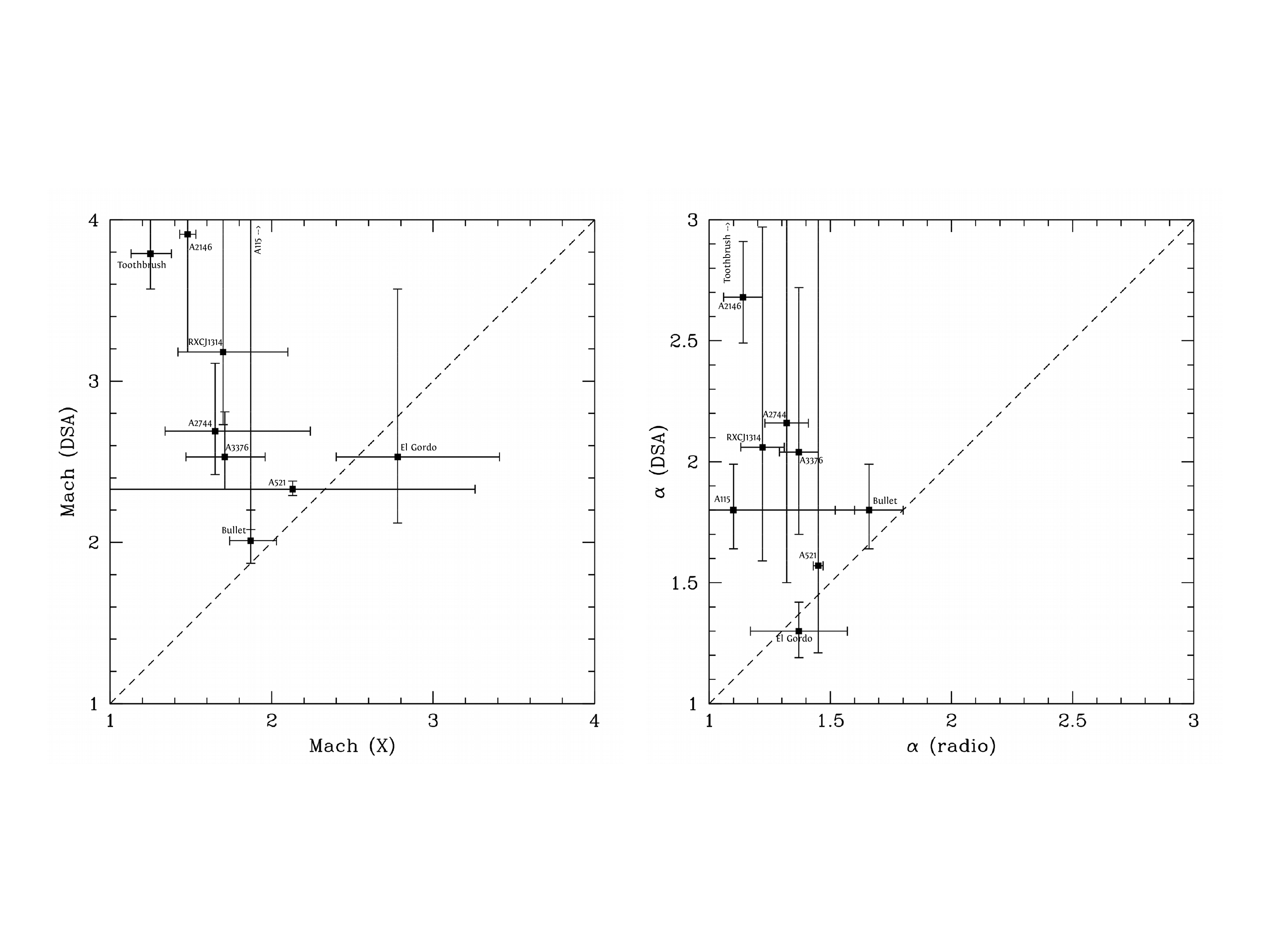}
 \caption{Observed Mach numbers and spectral indexes vs. expected values from DSA theory. The values used to produce the plots are those listed in Table~\ref{tab:x_vs_radio}. The dashed lines indicate the linear correlation as a reference.}
 \label{fig:mach_alpha}
\end{figure*}

\section{Results}\label{sec:results}

\subsection{Comparison between Mach numbers and spectral indexes}

In Table~\ref{tab:x_vs_radio} and Fig.~\ref{fig:mach_alpha} we compare the Mach number measured from X-ray observations ($\mach_X$) and the relic spectral index estimated from radio observations ($\alpha_{radio}$) with the expectations from DSA theory derived from Eq.~\ref{eq:alpha-mach} (where $\mach_X$ and $\alpha_{radio}$ are used to derive $\alpha_{DSA}$ and $\mach_{DSA}$, respectively) for the relics in our sample. In the cases of El Gordo, A521, and the Bullet Cluster, the spectral indexes are consistent. However, there is a discrepancy between the observed spectral index of the relic and that implied by DSA  in the majority of cases, confirming the findings of previous studies that showed that the Mach numbers derived from radio observations under the assumption of DSA are generally biased high in comparison to those coming from X-ray data \citep[e.g.,][]{akamatsu17a2255, urdampilleta18}. The inconsistency between radio and X-ray spectra is a long-standing problem of radio relics \citep[e.g.,][]{brunetti14rev}. However, such tension might be understood by looking at numerical simulations which show that the inconsistency between radio- and X-ray-derived Mach numbers might emerge from projection effects of multiple shock surfaces \citep{skillman13, hong15}. Furthermore, modifications to the basic DSA theory (e.g.,\ considering Alfv\'{e}nic drift or including superdiffusion regimes at the shocks, \citealt{kang18alfvenic, zimbardo18}) change the expected value of the spectral index of the accelerated particles from DSA predictions. 

\subsection{Efficiencies}

We calculate the acceleration efficiency of electrons $\eta_e$ and the relevant parameters that are necessary to reproduce the bolometric synchrotron luminosity in the sample of relics presented in Section~\ref{sec:sample} as a function of the magnetic field $B$. From Eq.~\ref{eq:luminosity}, the electron acceleration efficiency is

\begin{equation}\label{eq:eta_e}
 \eta_e \simeq \frac{2 \int_{\nu_0} L(\nu)\,{\rm d}\nu}{A \rho_u V_{sh}^3} \left( 1 - \frac{1}{\compr^2} \right)^{-1} \frac{1}{\Psi} {\frac{B_{cmb}^2 + B^2}{B^2}} 
,\end{equation}

\noindent
where $\Psi$ is given in Eq.~\ref{eq:psi} and depends on $p_{min}$. Since the Mach numbers observed in the X-rays and those predicted by DSA from the relic spectrum may be different (Fig.~\ref{fig:mach_alpha}), we explore three approaches. 

\begin{figure*}[t]
 \centering
 \includegraphics[width=.24\textwidth,trim={0cm 0cm 5cm 0.8cm},clip]{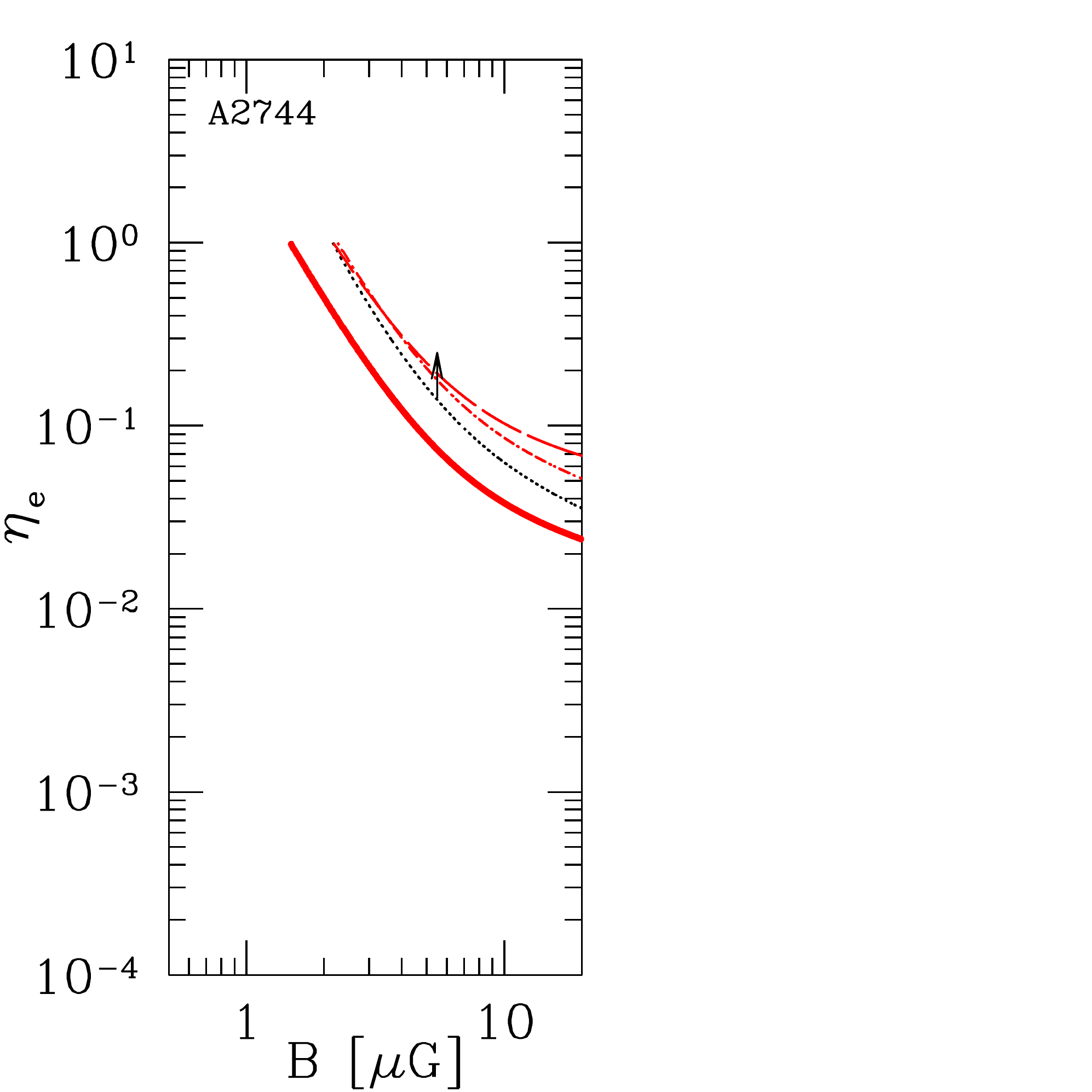}\hspace{-10mm}
 \includegraphics[width=.24\textwidth,trim={0cm 0cm 5cm 0.8cm},clip]{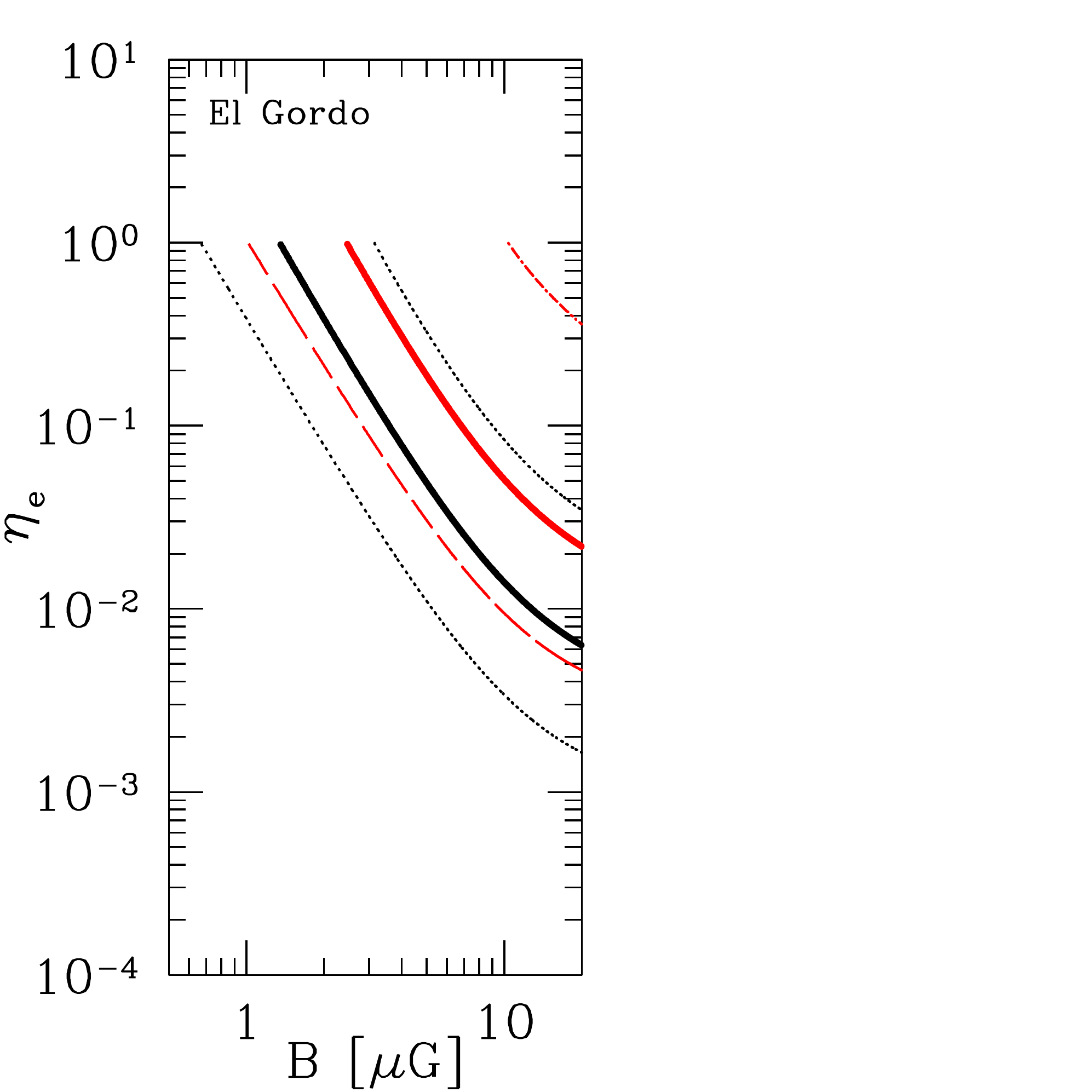}\hspace{-10mm}
 \includegraphics[width=.24\textwidth,trim={0cm 0cm 5cm 0.8cm},clip]{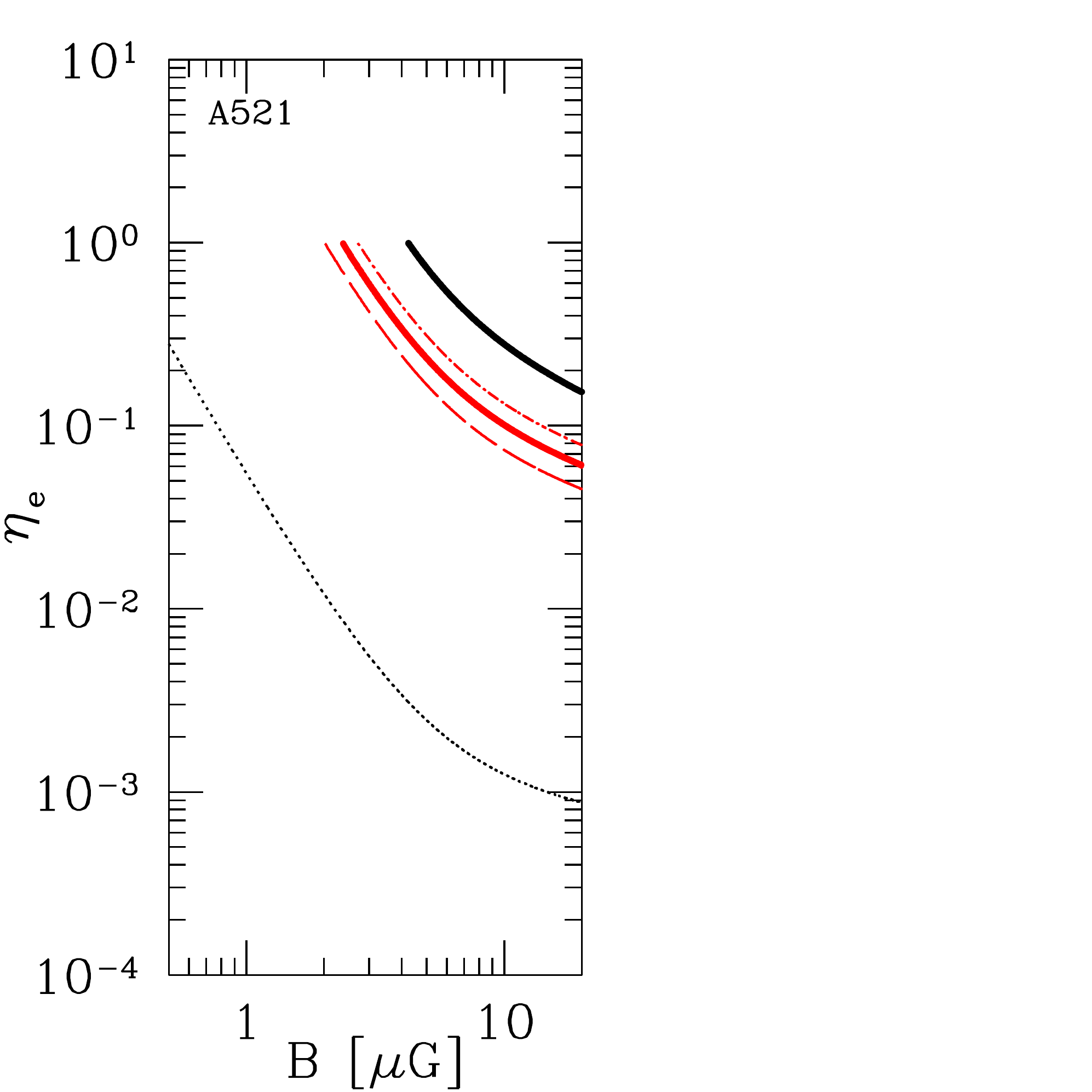}\hspace{-10mm}
 \includegraphics[width=.24\textwidth,trim={0cm 0cm 5cm 0.8cm},clip]{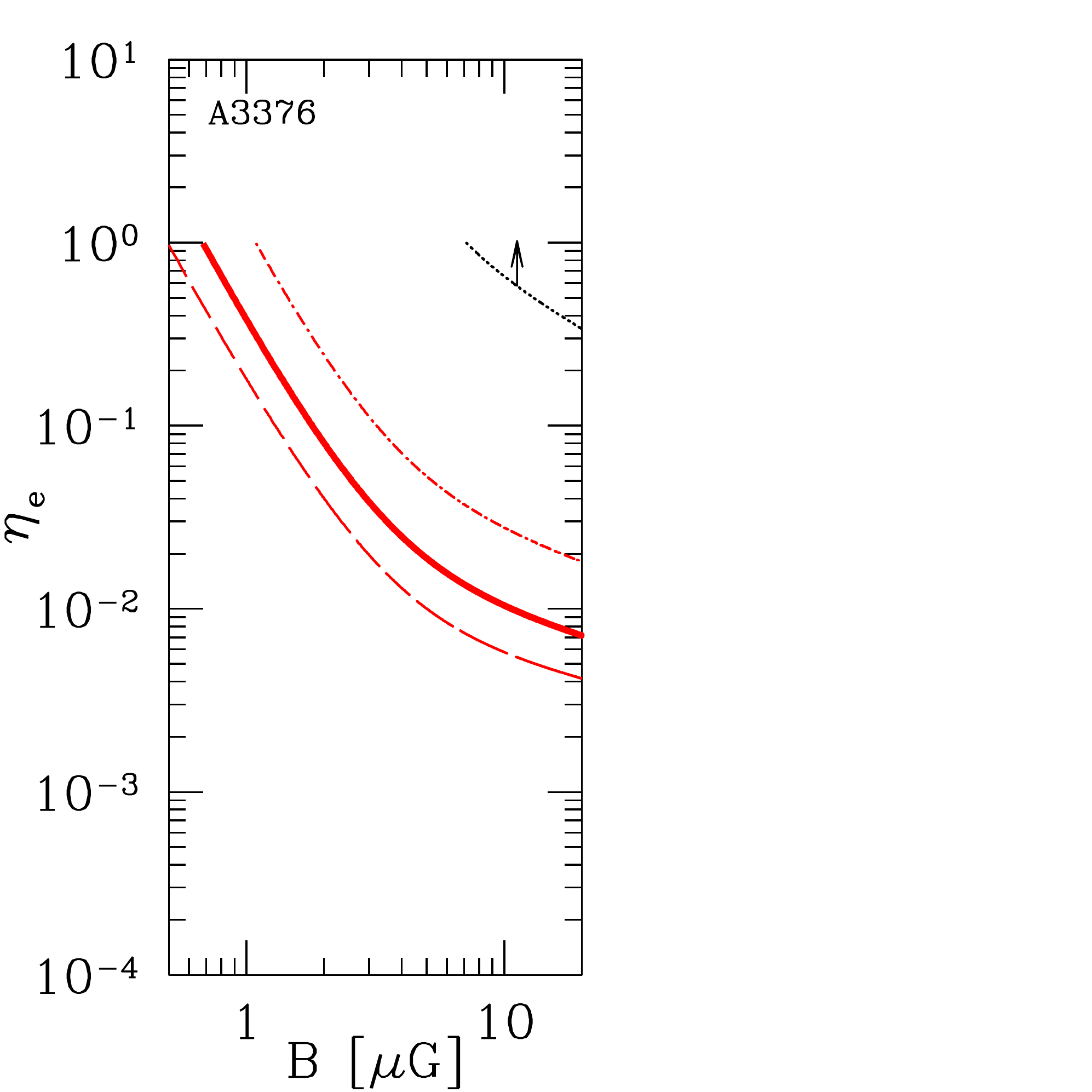} 
 \hspace*{-10mm} \\
 \vspace{5mm}
 \includegraphics[width=.24\textwidth,trim={0cm 0cm 5cm 0.8cm},clip]{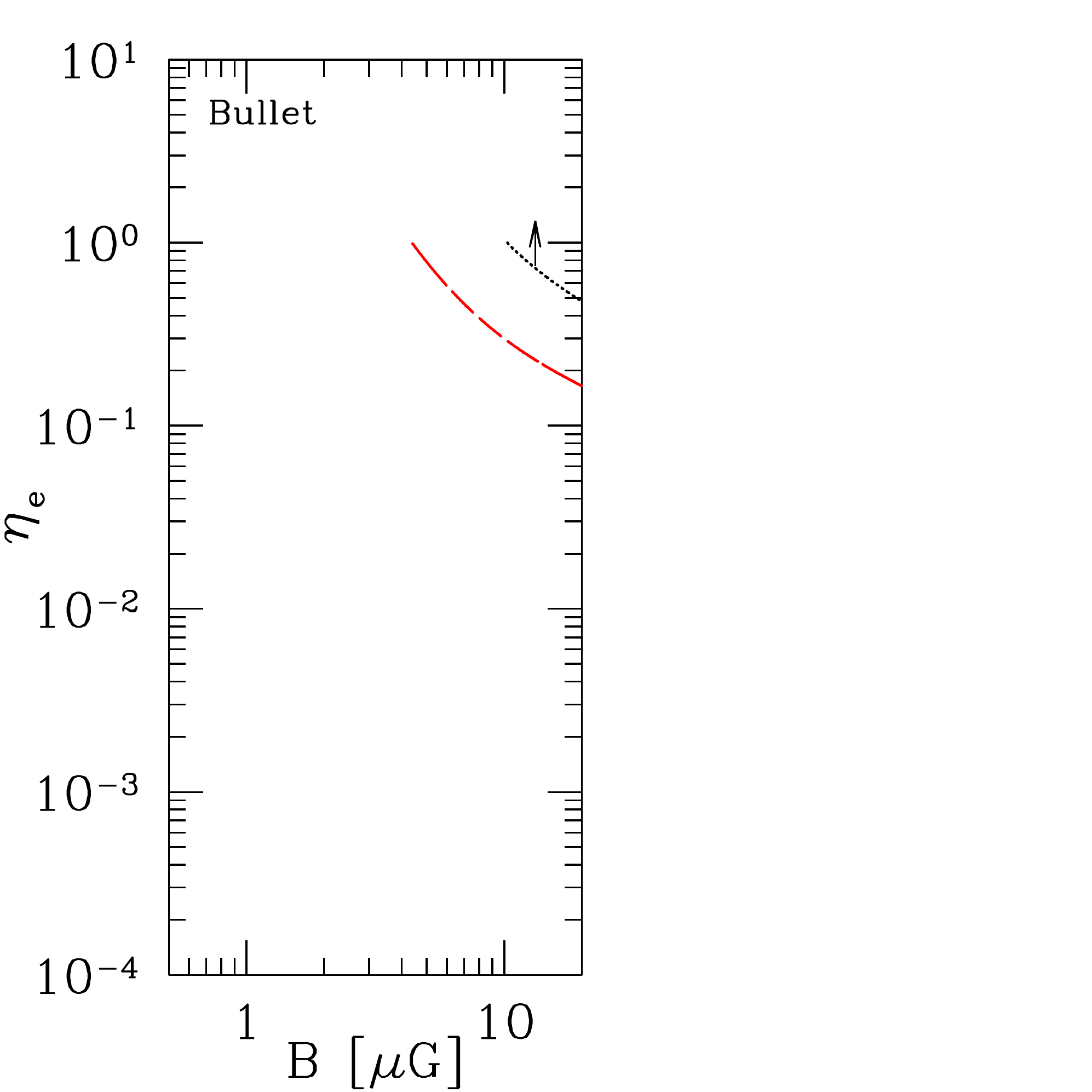}\hspace{-10mm} 
 \includegraphics[width=.24\textwidth,trim={0cm 0cm 5cm 0.8cm},clip]{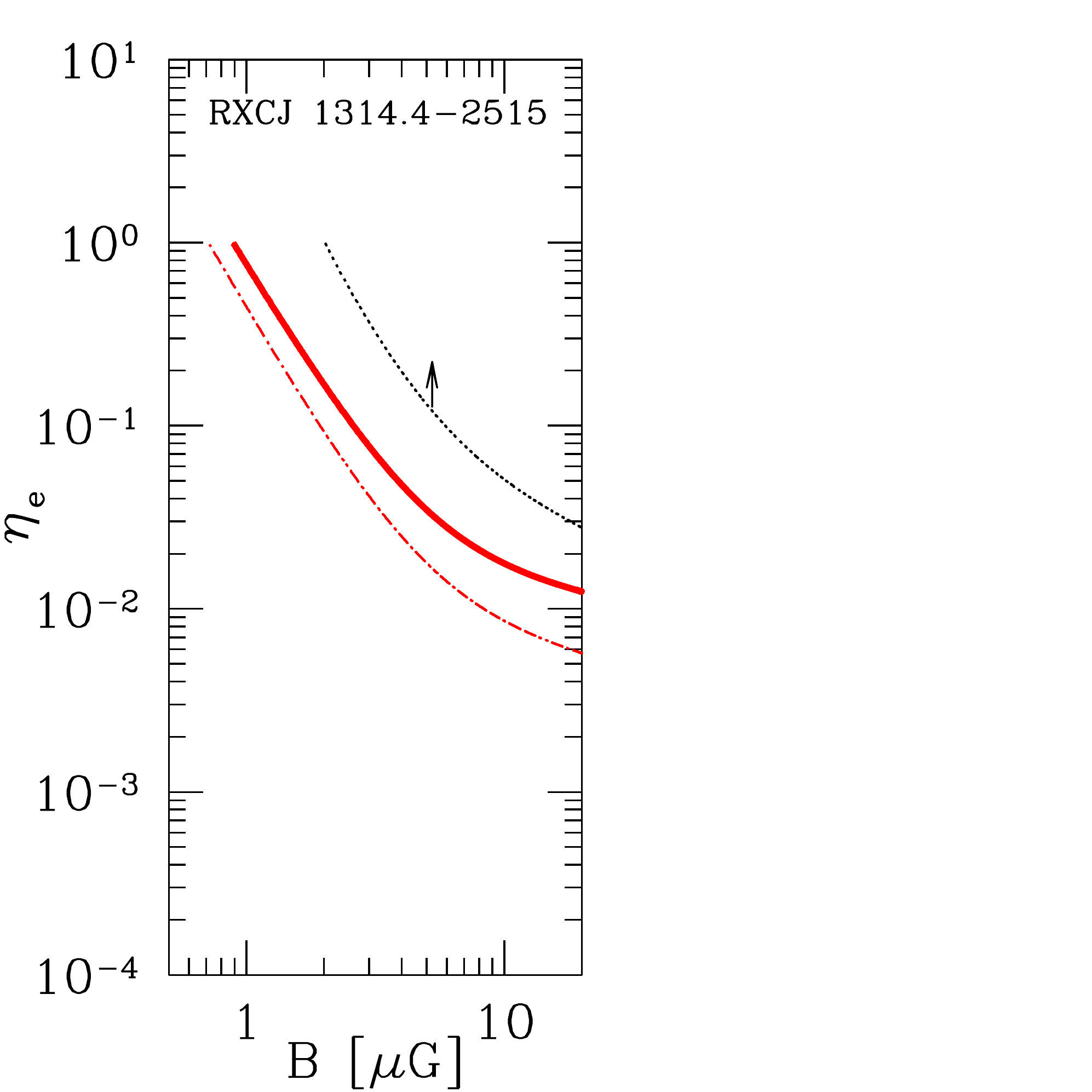}\hspace{-10mm}
 \includegraphics[width=.24\textwidth,trim={0cm 0cm 5cm 0.8cm},clip]{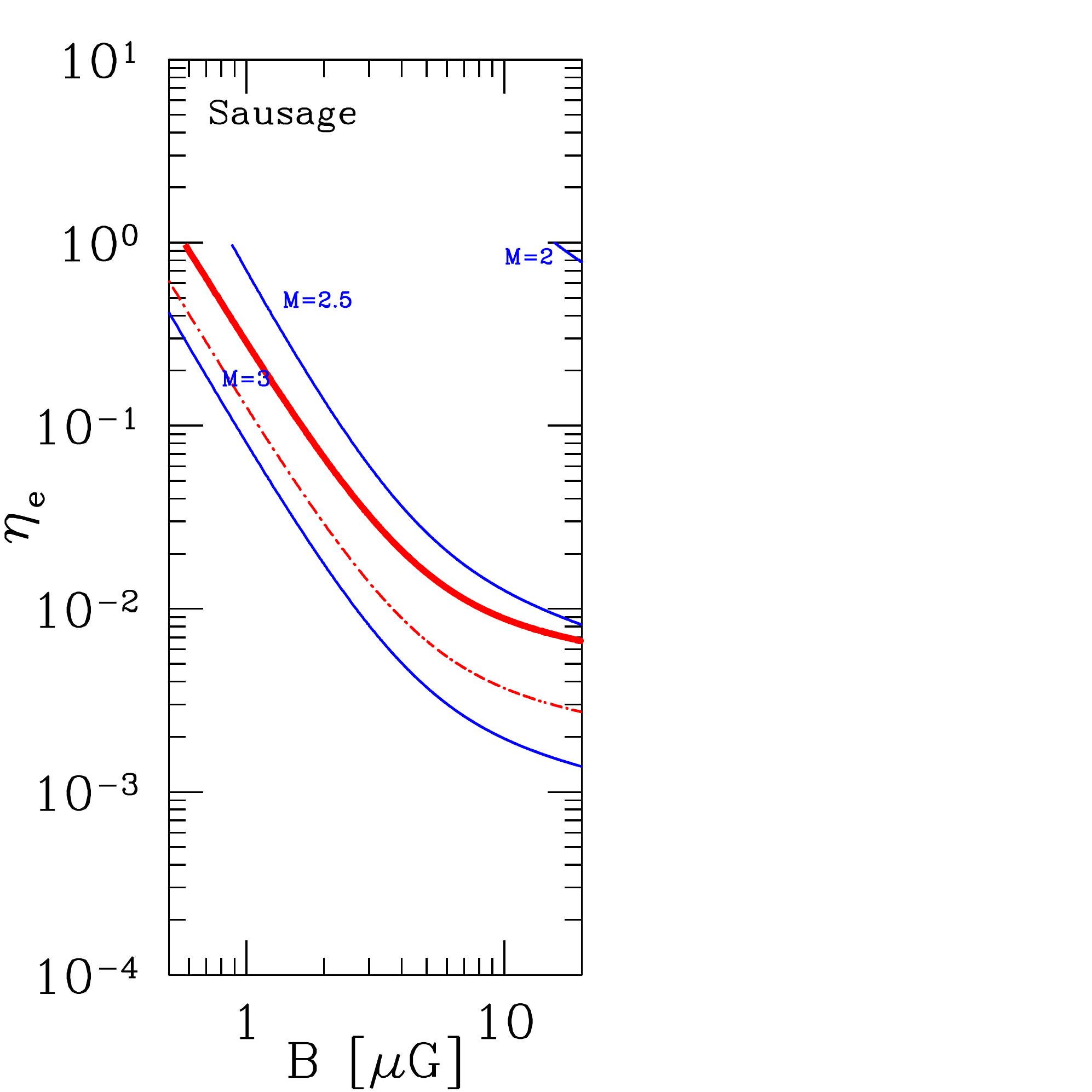}
 \hspace{-10mm}
 \caption{Electron acceleration efficiency for the radio relics of the sample vs. magnetic field in the downstream region. Calculations were performed using the Mach numbers listed in Table~\ref{tab:x_vs_radio} measured in the X-rays (\textit{black}) and derived from the integrated spectral index in the case of DSA (\textit{red}). Lines denote the best fit Mach numbers (\textit{solid}), the X-ray upper and lower bounds (\textit{dotted}), and the upper (\textit{long dashed}) and lower (\textit{dot-dashed}) bounds of $\mach_{DSA}$. For the Sausage relic, lines in \textit{blue} represent assumed Mach numbers of $\mach = 2, 2.5,$ and  $3$ (from top to bottom).}
 \label{fig:eff_total}
\end{figure*}

In the first two approaches, we assume standard DSA and use Eqs.~\ref{eq:delta-inj} and \ref{eq:alpha-mach} to connect the Mach number and the radio spectrum. The two approaches differ for the values of \mach\ that are assumed. In the first one (Section~\ref{sec:xray_mach}), we use $\mach_X$ to calculate the spectrum of the accelerated particles. This approach is therefore based on X-ray observations. This scenario is  at least relevant for those relics where the radio- and X-ray-derived Mach numbers are consistent (Fig.~\ref{fig:mach_alpha}). In the second approach (Section~\ref{sec:radio_mach}), we assume that the average Mach number of the shocks is given by $\mach_X$ (and we therefore use $\mach_{X}$ to get the energy flux at the shock; Eq.~\ref{eq:conservation}) and that the acceleration of electrons that generate radio relics is dominated by the sites in the shock surfaces where the Mach numbers $\ge \mach_{DSA}$. This approach is motivated by numerical simulations \citep{hong15, roh19} and might explain both radio and X-ray observations. In the third approach (Section~\ref{sec:beyonddsa}), we go beyond the standard DSA model. We assume that a shock with Mach number $\mach_X$ accelerates particles with a spectrum of $\delta= 2 \alpha_{radio}$, where $\alpha_{radio}$ is the observed radio spectrum. This scenario allows us to match both radio and X-ray observations and is motivated by recent theoretical proposals \citep[e.g.,][]{zimbardo18}. In all these approaches, we use the X-ray Mach numbers $\mach_X$ estimated from the surface brightness analysis to calculate the energy flux at the shock surface (Eq.~\ref{eq:conservation}) because they are better constrained than those obtained with the spectral analysis.

\subsubsection{Mach numbers measured in the X-rays}\label{sec:xray_mach}

\begin{figure}[t]
 \centering
 \includegraphics[width=\hsize,trim={0cm 0cm 0cm 0cm},clip]{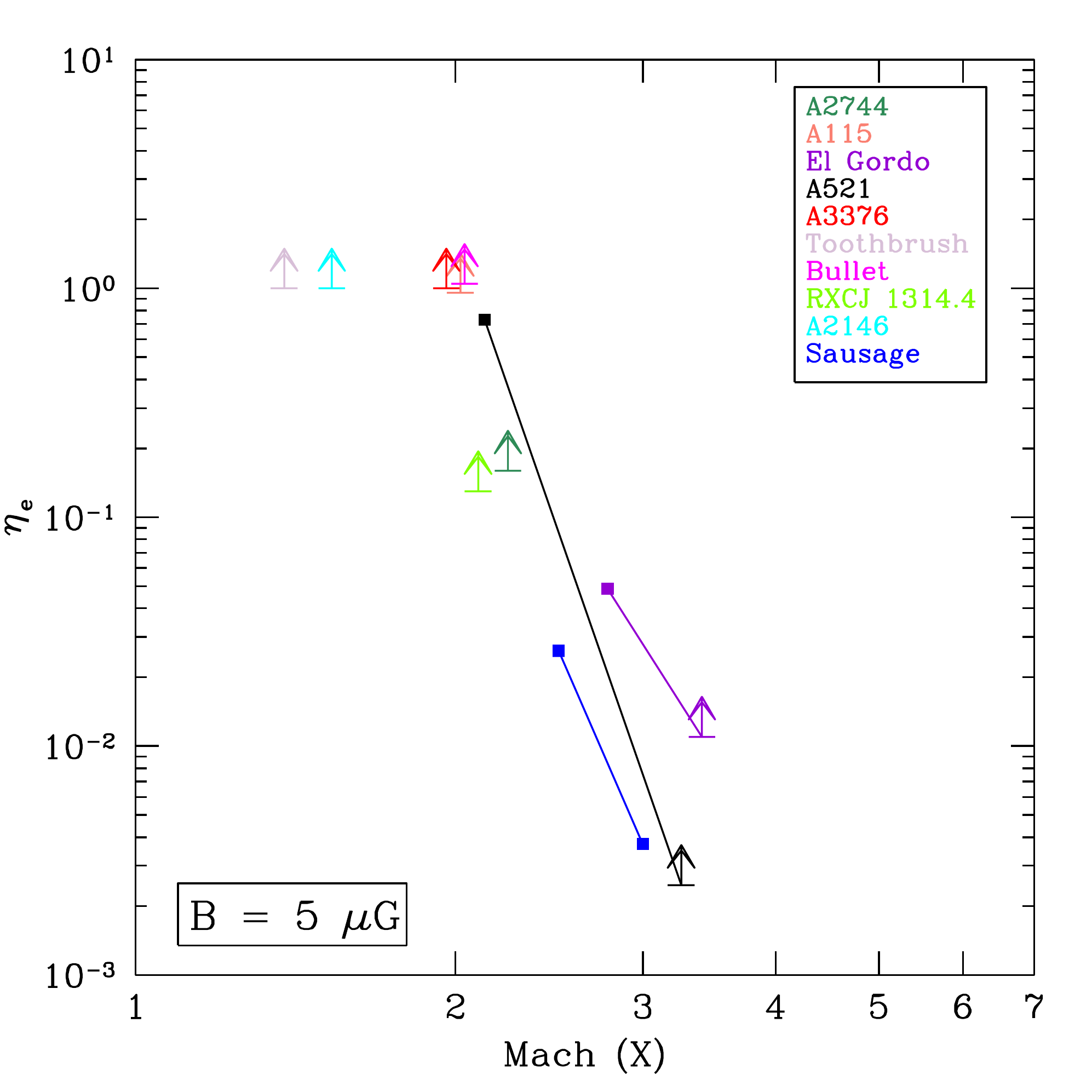}
 \caption{Electron acceleration efficiency vs. $\mach_X$ obtained for $B = 5$ \muG. Most of the relics are reported as lower limits because it was only possible to compute the acceleration efficiency for the upper bounds of $\mach_X$. For the Sausage relic, we assumed $\mach_X = 2.5$ and 3 as no shock is observed in surface brightness.}
 \label{fig:B_eta_computed_xray}
\end{figure}

\begin{figure}[t]
 \centering
 \includegraphics[width=\hsize,trim={3.8cm 0.5cm 3.8cm 1.2cm},clip]{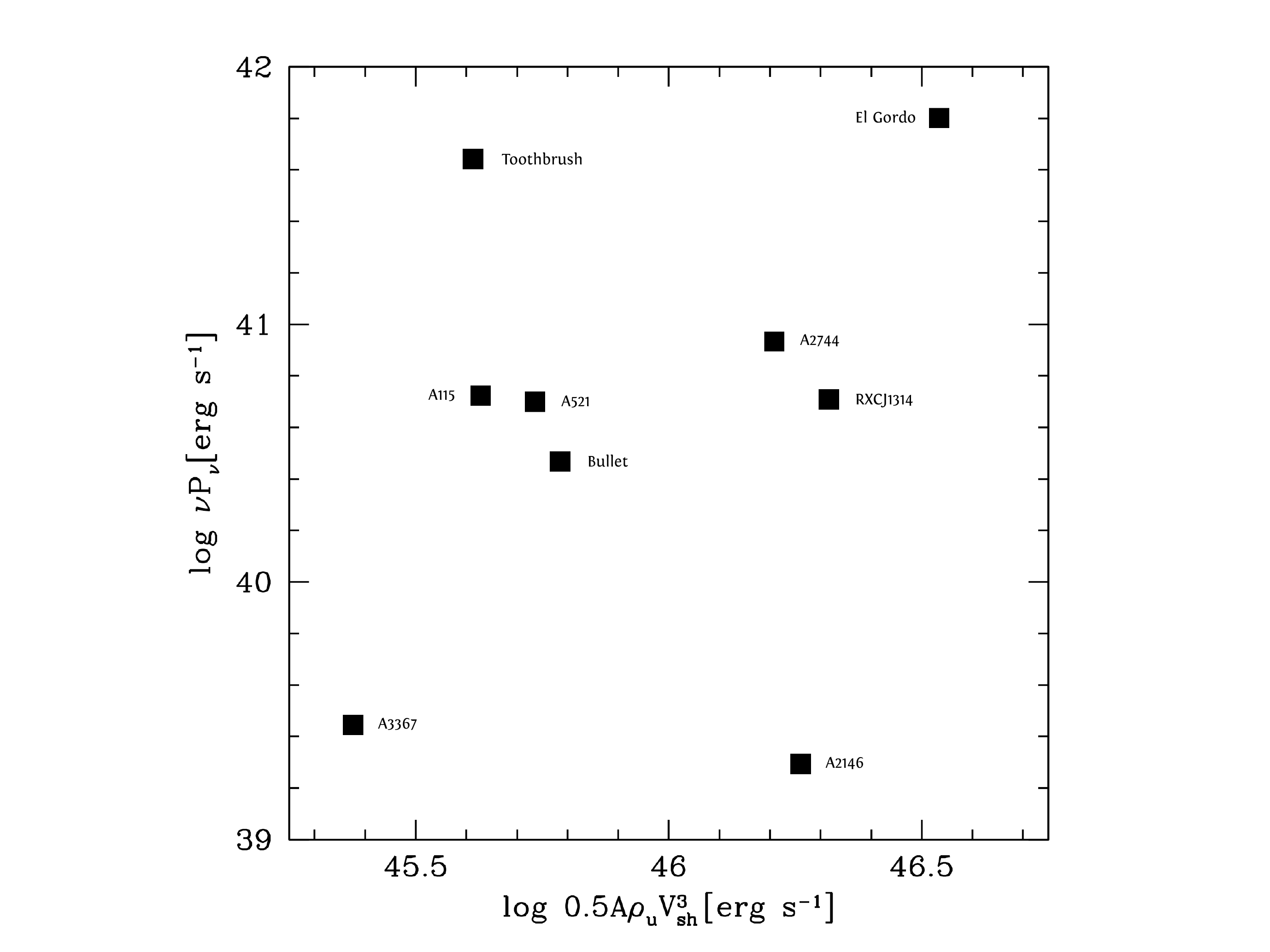}
 \caption{Kinetic energy flux through shock surface vs. relic radio power. The values reported in Table~\ref{tab:efficiency_param} were used to compute the two quantities.}
 \label{fig:kineticflux}
\end{figure}

In Fig.~\ref{fig:eff_total} we show the electron acceleration efficiency versus magnetic field for the relics in the sample using the Mach number measured in the X-rays (black lines). We note that in most cases we report only a lower limit on $\eta_e$ in the $(B,\eta_e)$ plane because the observed radio luminosity cannot be matched even if the shock were to accelerate the entire distribution of thermal electrons (namely $p_{min} < p_{th}$ in Eq.~\ref{eq:p_min}, \cf\ Fig.~\ref{fig:definition}). For A115, Toothbrush cluster, and A2146, no solution is found in Eq.~\ref{eq:p_min} because the Mach numbers measured in the X-rays are too low and for this reason they are not reported in Fig.~\ref{fig:eff_total}. \\
\indent
The acceleration efficiency of CRp for weak shocks is likely $<1\%$ \citep[e.g.,][]{ryu19}, in order for $\gamma$-ray emission from clusters to be consistent with the \fermi-LAT upper limits \citep[e.g.,][]{vazza16, ha19arx}. Although the ratio of CRe/CRp acceleration is not measured for weak ICM shocks, we can reasonably assume that, as in the case of SNRs, the great majority of the energy flux at these shocks goes into CRp. For this reason, in the following we consider a conservative value of $\eta_e < 0.1$, which is generally associated to (protons in) strong SNR shocks ($\mach \sim 10^3$), and a more realistic value of $\eta_e < 0.01$. In Fig.~\ref{fig:B_eta_computed_xray} we compute the acceleration efficiency that is requested to match the observed radio luminosity of the relics as a function of the shock Mach number measured in the X-rays at fixed downstream magnetic field of $B = 5$ \muG; smaller magnetic fields will increase the requested value of $\eta_e$ (see Eq.~\ref{eq:eta_e}). For the Sausage relic, where no surface brightness jump is observed in the X-rays, we assume $\mach_X = 2.5$ and $\mach_X = 3$. For all the relics except A521 and El Gordo, it is possible to compute (\ie,\ $p_{min} > p_{th}$ in Eq.~\ref{eq:p_min}) the efficiency using  only the upper bounds on the shock Mach number, and for this reason lower limits on the efficiency are reported in Fig.~\ref{fig:B_eta_computed_xray}. \\
\indent
Looking at the distribution of acceleration efficiencies in Figs.~\ref{fig:eff_total} and \ref{fig:B_eta_computed_xray}, the model of DSA of thermal electrons is ruled out for the great majority of our relics. Even if we adopt an optimistic threshold of $\eta_e<0.1$, the only two relics whose bolometric radio luminosities can be reproduced by DSA from the thermal pool are those in El Gordo and A521, which are also those where $\mach_X$ and $\mach_{DSA}$ are consistent (Fig.~\ref{fig:mach_alpha}); we note that this is only possible for A521 because of the large uncertainty of $\mach_{X}$, leading to an upper bound value which is significantly larger than $\mach_{DSA}$. In general, as already mentioned at the end of Section~\ref{sec:computation}, Fig.~\ref{fig:B_eta_computed_xray} clearly shows the importance of the Mach number in the acceleration process: only $\mach \gtrsim 2.5$ have an energy flux and produce an accelerated spectrum that is sufficient to explain the luminosity of radio relics with the DSA of thermal ICM electrons. \\
\indent
The origin of radio relics can also be investigated by looking at possible correlations between the properties of shocks and the associated radio luminosity. \citet{colafrancesco17} studied a sample of radio relics and investigated the connection between relic radio power and shock Mach number. They interpreted the lack of correlation between these quantities as evidence against an origin of relics via DSA of thermal electrons. However, in general, the shock Mach number is not indicative of the energetics of the shock and $\mach$ is only a fair measure of shock speed if the upstream temperature is the same for all the relics in a sample, and this is not the case. 
For this reason, we consider the kinetic energy flux across the surface of the shock, that is $1/2 A \rho_u V_{sh}^3$, which can be measured by X-ray observations (that provide upstream density, temperature, and Mach number) and radio observations (that provide the shock/relic surface area). We compare the radio luminosity and kinetic energy flux at the shocks for our radio relics but we do not find any clear correlation (Fig.~\ref{fig:kineticflux}). For similar kinetic energy flux at the shocks, we notice that the radio luminosity of relics differs by more than two orders of magnitude. This further suggests that DSA of thermal electrons is not the general mechanism acting for radio relics.

\subsubsection{Matching X-ray and DSA Mach numbers with simulations}\label{sec:radio_mach}

\begin{figure}[t]
 \centering
 \includegraphics[width=\hsize,trim={0cm 0cm 0cm 0cm},clip]{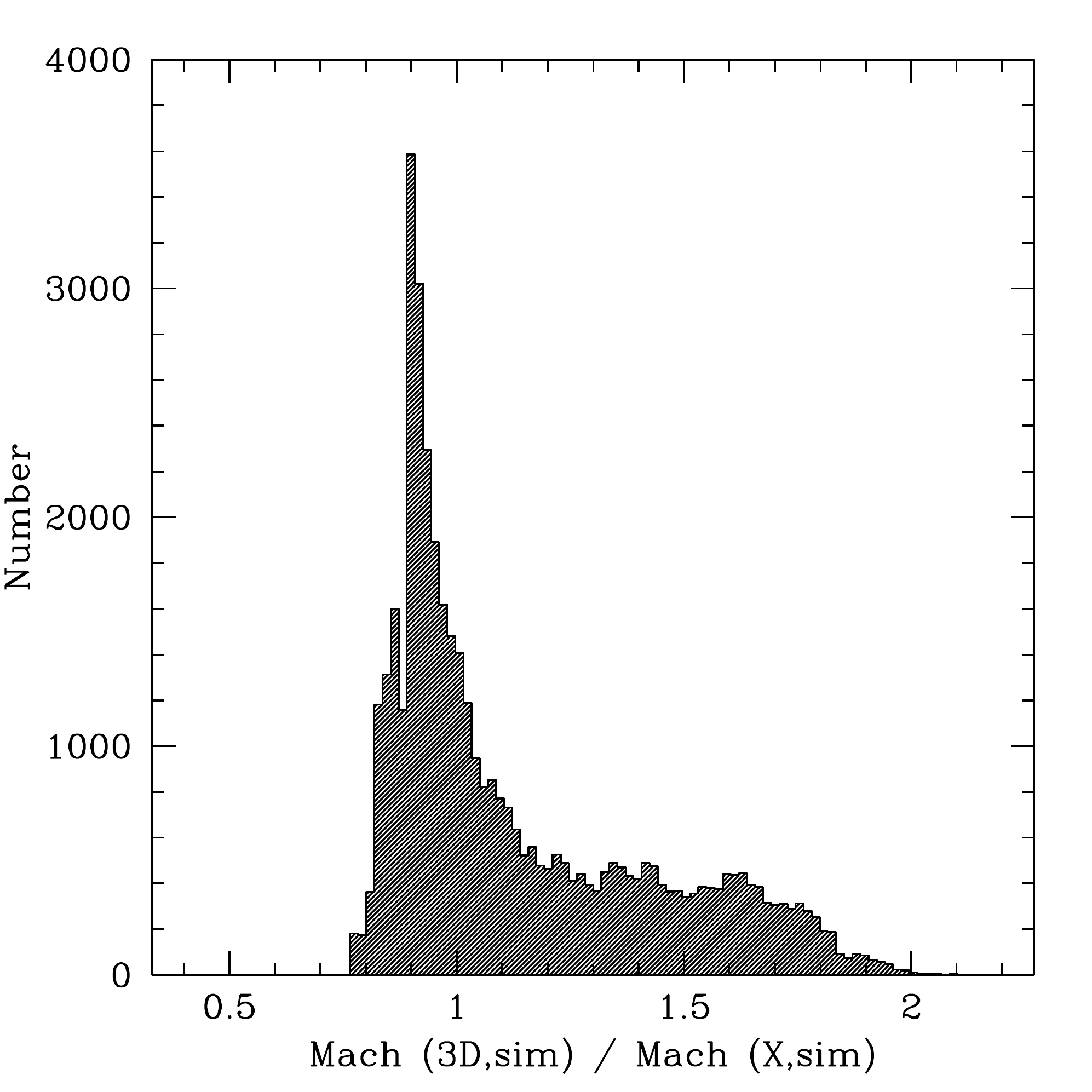}
 \caption{Distribution of the ratio between the 3D Mach number $\mach_{3D,sim}$ and the average X-ray Mach number $\mach_{X,sim}$ for five radio relics extracted from our simulations.}
 \label{fig:cumula_xray}
\end{figure}

Numerical simulations show that Mach numbers implied by DSA from the radio spectra of relics in simulations are slightly larger than Mach numbers measured in the X-rays \citep[e.g.,][]{hong15} because in simulations the Mach number is not constant across the shock surface and consequently the synchrotron emission from accelerated electrons is mainly contributed by regions where the Mach number is larger. For this reason, in this section we compute the electron acceleration efficiency for our relics by assuming the spectrum of the accelerated electrons and the Mach number implied by DSA from radio observations (Eq.~\ref{eq:alpha-mach}). We use Eq.~\ref{eq:eta_e}, but in this case the surface $A$ is the area covered by large Mach numbers, $\geq \mach_{DSA}$. \\
\indent
The fraction $f_{DSA}$ of the shock area that is covered by Mach numbers $\geq \mach_{DSA}$ is estimated with the simulations described in \citet{roh19}. These simulations followed turbulence, magnetic fields, and shocks in the ICM using a model cluster in a controlled box to achieve a high resolution. With the data of the highest resolution (the size of grid zone $\Delta x = 3.9$ kpc) simulation, a number of mock radio relics are identified. We extract about $ 43000$ grid zones, which together cover five shock surfaces of radio relics, and calculate the 3D Mach number of these cells, $\mathcal{M}_{3D,sim}$ \citep[see][]{roh19}. In Fig.~\ref{fig:cumula_xray} we show the distribution of $\mathcal{M}_{3D,sim}$ normalized to the average Mach number of the shock surface to which the cells belong, $\mathcal{M}_{X,sim}$. Following \citet{hong15}, $\mathcal{M}_{X,sim}$ is calculated from the temperature jump projected along the line of sight that is measured in simulations across the shock surfaces. In line with \citet{hong15}, we find that shocked cells with a Mach number that is approximately equal to or greater than two times the X-ray Mach number of the shock are extremely rare. We determined $f_{DSA}$ for our relics from the fraction of shocked cells with $\mathcal{M}_{3D,sim} / \mathcal{M}_{X,sim} \geq \mach_{DSA}/\mach_X$; the ratio $\mach_{DSA}/\mach_X$ is derived from the values of $\mach_{DSA}$ (with its error bounds) and the central value of $\mach_X$ reported in Table~\ref{tab:x_vs_radio}. The resulting $f_{DSA}$ are reported in Table~\ref{tab:fraction}. 

\begin{table}[t]
 \centering
 \caption{Values of $f_{DSA}$ taking into accounts the error bounds of $\mach_{DSA}$ and the best fit of $\mach_X$ for the relics in our sample. For the Sausage Cluster we assumed $\mach_X = 2.5$.}
 \label{tab:fraction}
  \begin{tabular}{lccc} 
  \hline
  Cluster name & $f_{DSA}^{-}$ & $f_{DSA}^{}$ & $f_{DSA}^{+}$ \\
  \hline
  A2744 & 0.1783 & 0.0985 & 0.0099 \\
  El Gordo & 1.000 & 0.7788 & 0.2901 \\
  A521 & 0.4392 & 0.4195 & 0.4016 \\
  A3376 & 0.2405 & 0.1699 & 0.0882 \\
  Bullet Cluster & 0.4202 & 0.4392 & 0.3450 \\
  RXC J1314.4-2515 & 0.1086 & 0.0117 & -- \\
  Sausage Cluster & 0.1535 & 0.0420 & -- \\
  \hline
  \end{tabular}
\end{table}

\begin{figure}[t]
 \centering
 \includegraphics[width=\hsize,trim={0cm 0cm 0cm 0cm},clip]{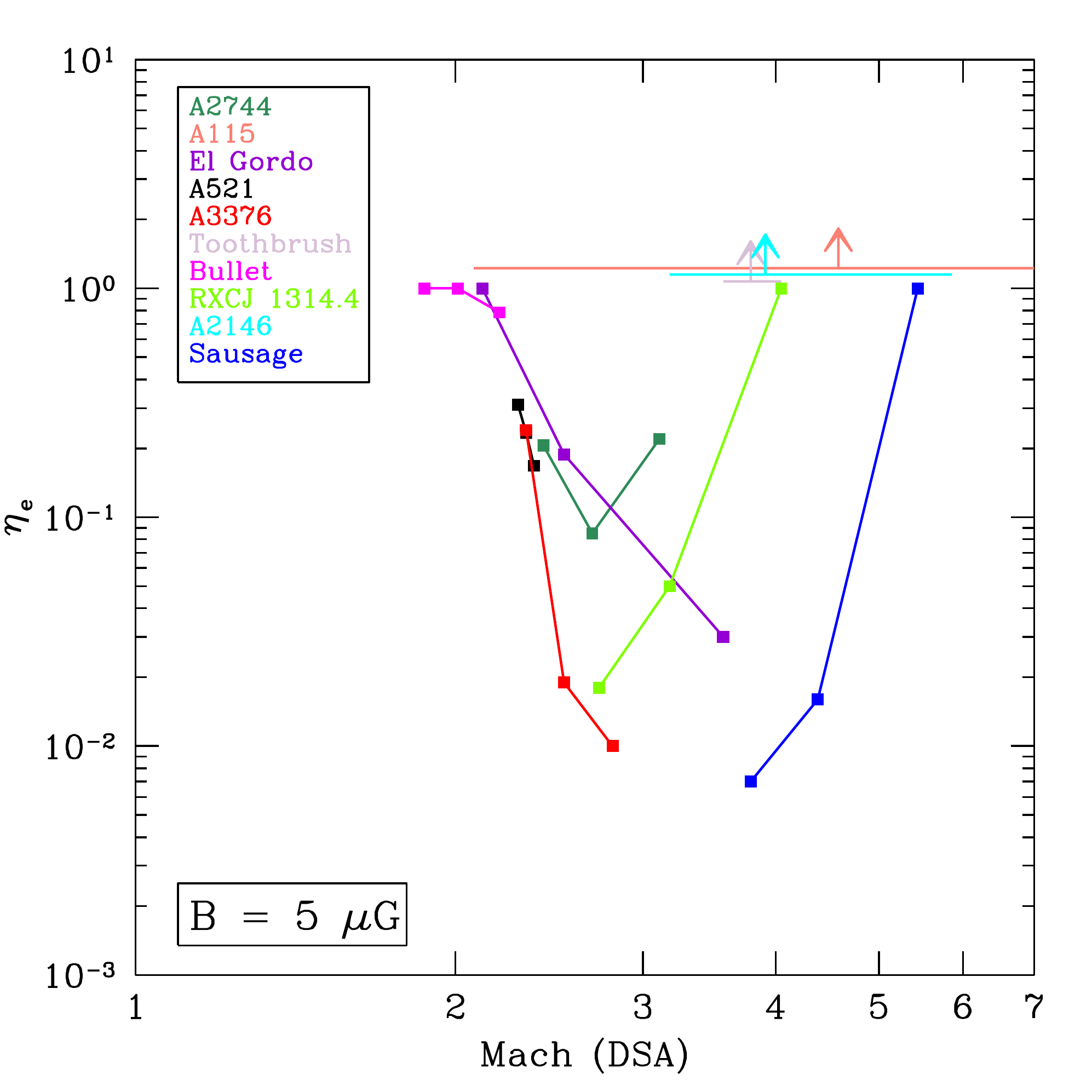}
 \caption{Electron acceleration efficiency vs. $\mach_{DSA}$ obtained for $B = 5$ \muG. For the Sausage relic, $f_{DSA}$ was calculated by assuming $\mach_X = 2.5$. Upper limits at $\eta_e>1$ are slightly offset for display purposes.}
 \label{fig:B_eta_computed_DSA}
\end{figure}

We use Eq.~\ref{eq:eta_e} to calculate the acceleration efficiency that is necessary to match the observed synchrotron luminosity assuming a surface $A' = A \times f_{DSA}$ and $\mach_{DSA}$ (\cf\ Tab.~\ref{tab:x_vs_radio}). The electron acceleration efficiency versus magnetic field for the relics in the sample computed using the Mach numbers implied by DSA from the relic integrated spectral indexes are shown in Fig.~\ref{fig:eff_total} (red lines). As before, in Fig.~\ref{fig:B_eta_computed_DSA} we summarize the results by showing the acceleration efficiency that is required to match the observed radio luminosity of the relics as a function of the shock Mach number derived under DSA assumption at a fixed downstream magnetic field of $B = 5$ \muG. \\
\indent
Even if we consider an optimistic acceleration efficiency of $\eta_e < 0.1$, we find that only four out of ten relics (namely: A2744, El Gordo, A3376, and RXCJ 1314.4-2515) can be explained via DSA assuming an optimistic value of $B > 5$ \muG. On the other hand, an efficiency $\eta_e < 0.01$ is obtained only for A3376 and only by considering $\mach_{DSA} = 3.18$ (\ie,\ its upper bound on $\mach_{DSA}$) and $B > 5$ \muG. Therefore, also in this case, we conclude that even if this approach matches the synchrotron spectral index and the average Mach number measured in the X-rays, it energetically fails to reproduce the synchrotron luminosity observed in the great majority of our radio relics. \\
\indent
We note that the correction factor and the efficiency obtained with this procedure are anchored to the Mach number measured in the X-rays for the relics (Table~\ref{tab:x_vs_radio}). Although there are possible systematic errors involved in measuring Mach numbers in the X-rays \citep[e.g.,][]{markevitch07rev}, we believe that these Mach numbers for our relics are relatively reliable for two main reasons: (i) the Mach numbers from density and temperature jumps (that have
very different dependencies on \mach) are generally consistent for our relics, and (ii) the presence of double relic systems in our sample suggests that at least in a large fraction of our relics the shocks propagate close to the plane of the sky, thus minimizing projection effects.

\subsubsection{Beyond standard DSA}\label{sec:beyonddsa}

\begin{figure*}[ht]
 \centering
 \includegraphics[width=.24\textwidth,trim={0cm 0cm 5cm 0.8cm},clip]{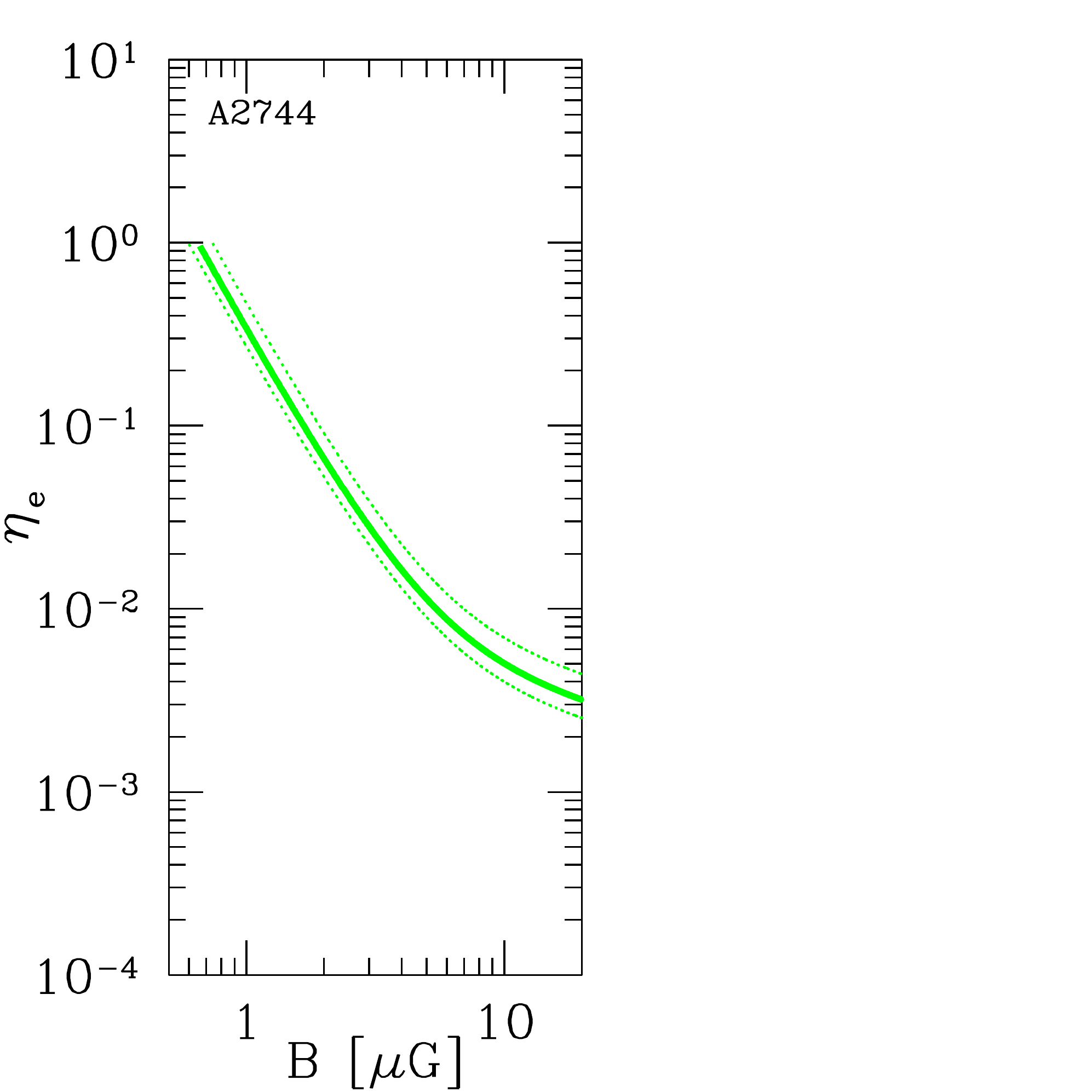}\hspace{-10mm}
 \includegraphics[width=.24\textwidth,trim={0cm 0cm 5cm 0.8cm},clip]{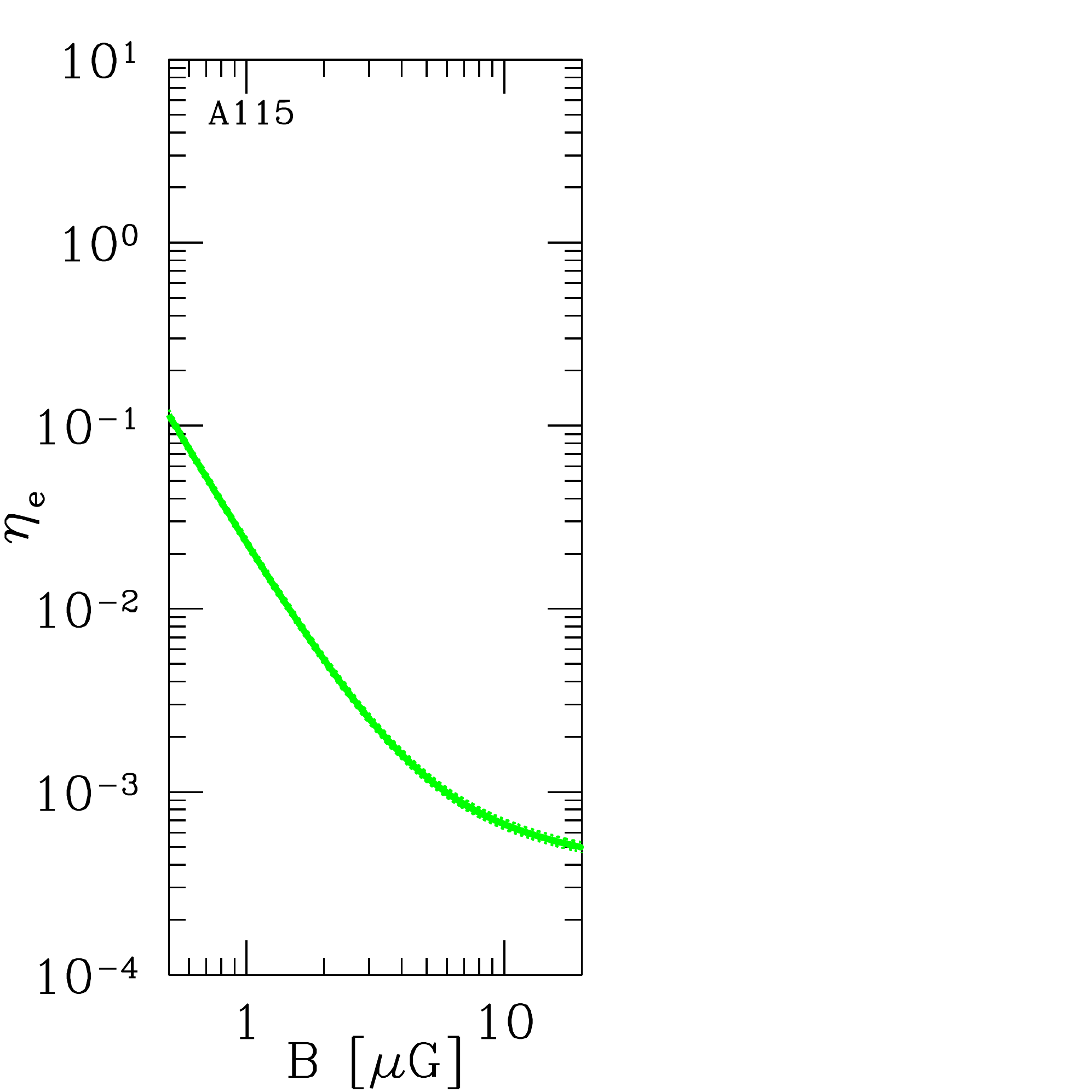}\hspace{-10mm} 
 \includegraphics[width=.24\textwidth,trim={0cm 0cm 5cm 0.8cm},clip]{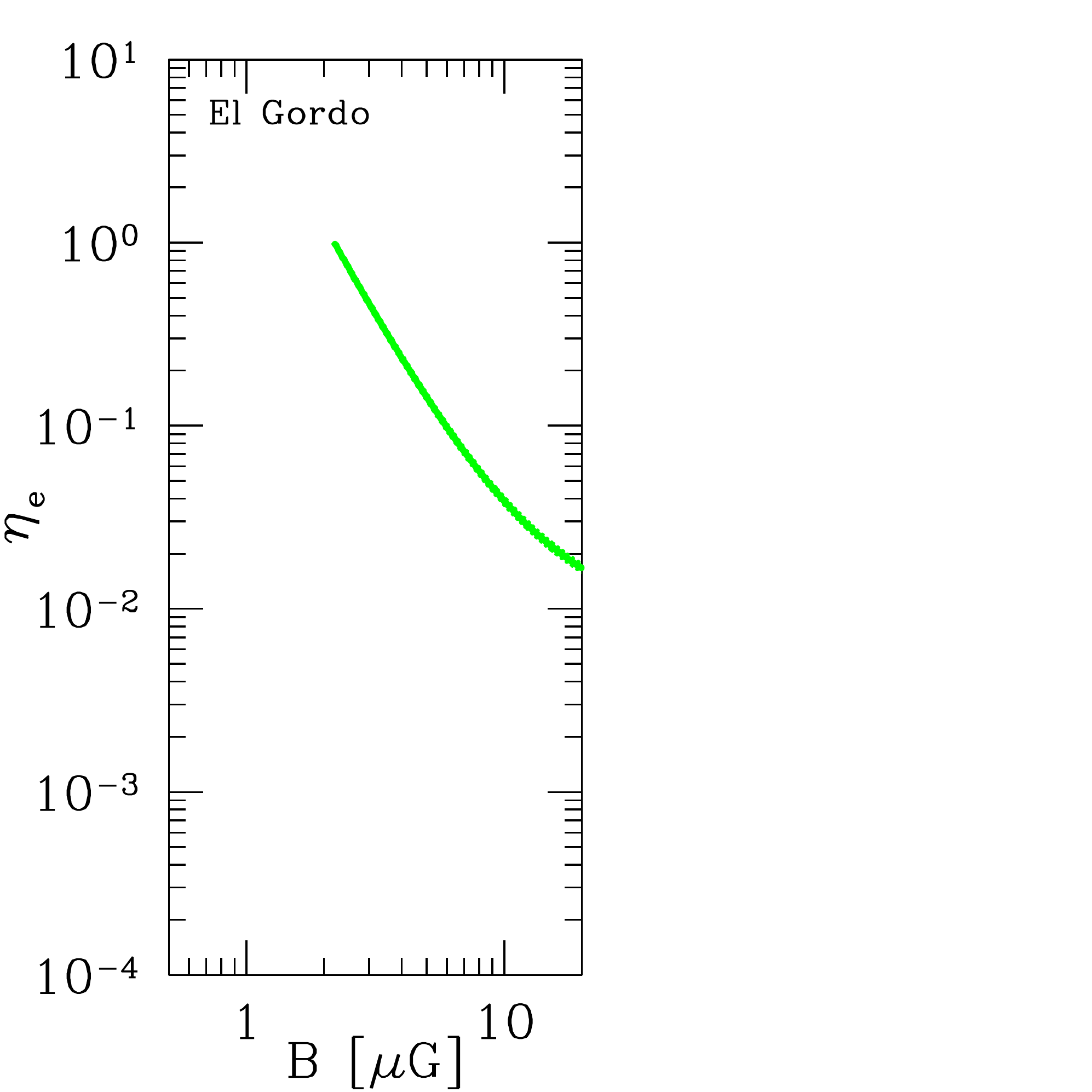}\hspace{-10mm}
 \includegraphics[width=.24\textwidth,trim={0cm 0cm 5cm 0.8cm},clip]{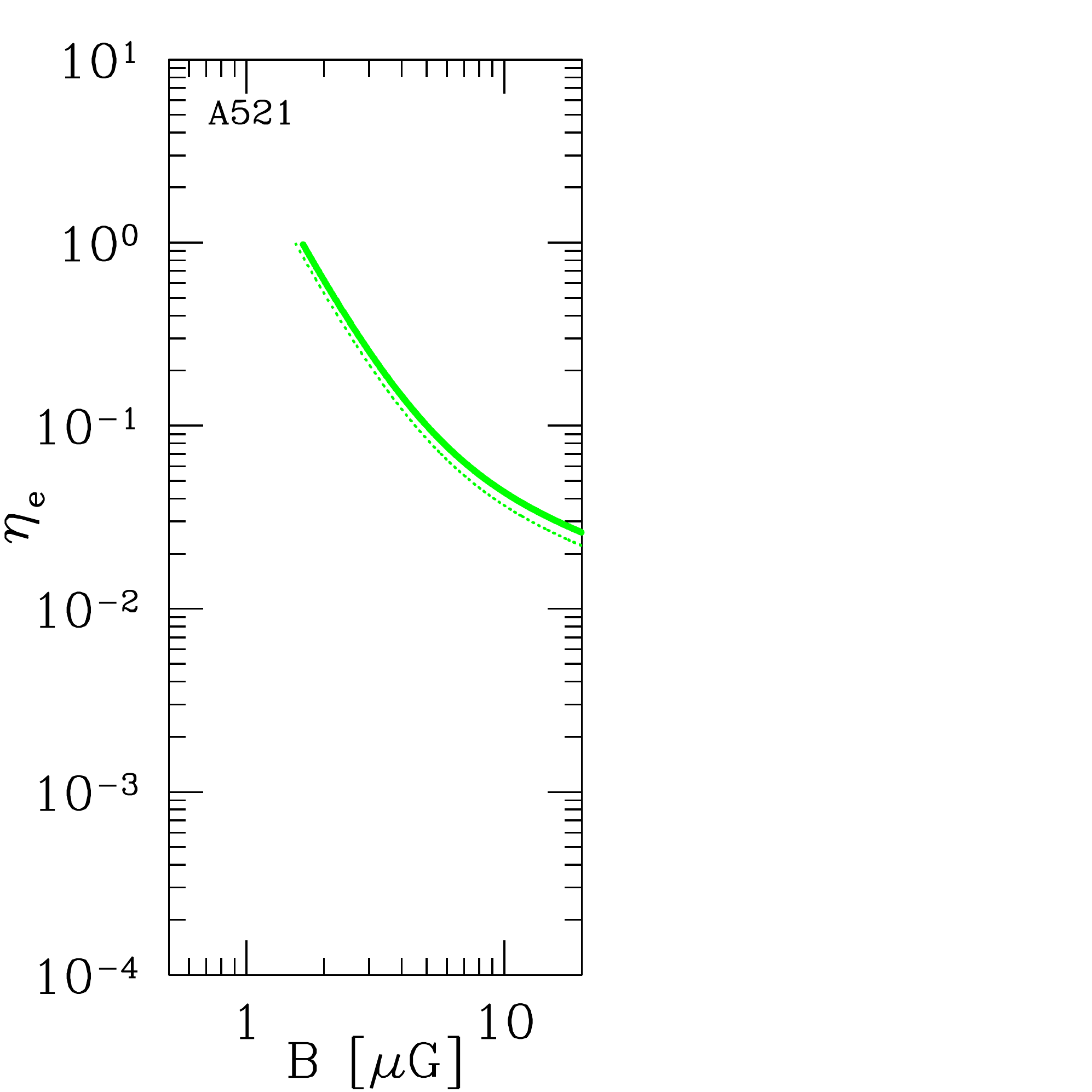}\hspace{-10mm}
 \includegraphics[width=.24\textwidth,trim={0cm 0cm 5cm 0.8cm},clip]{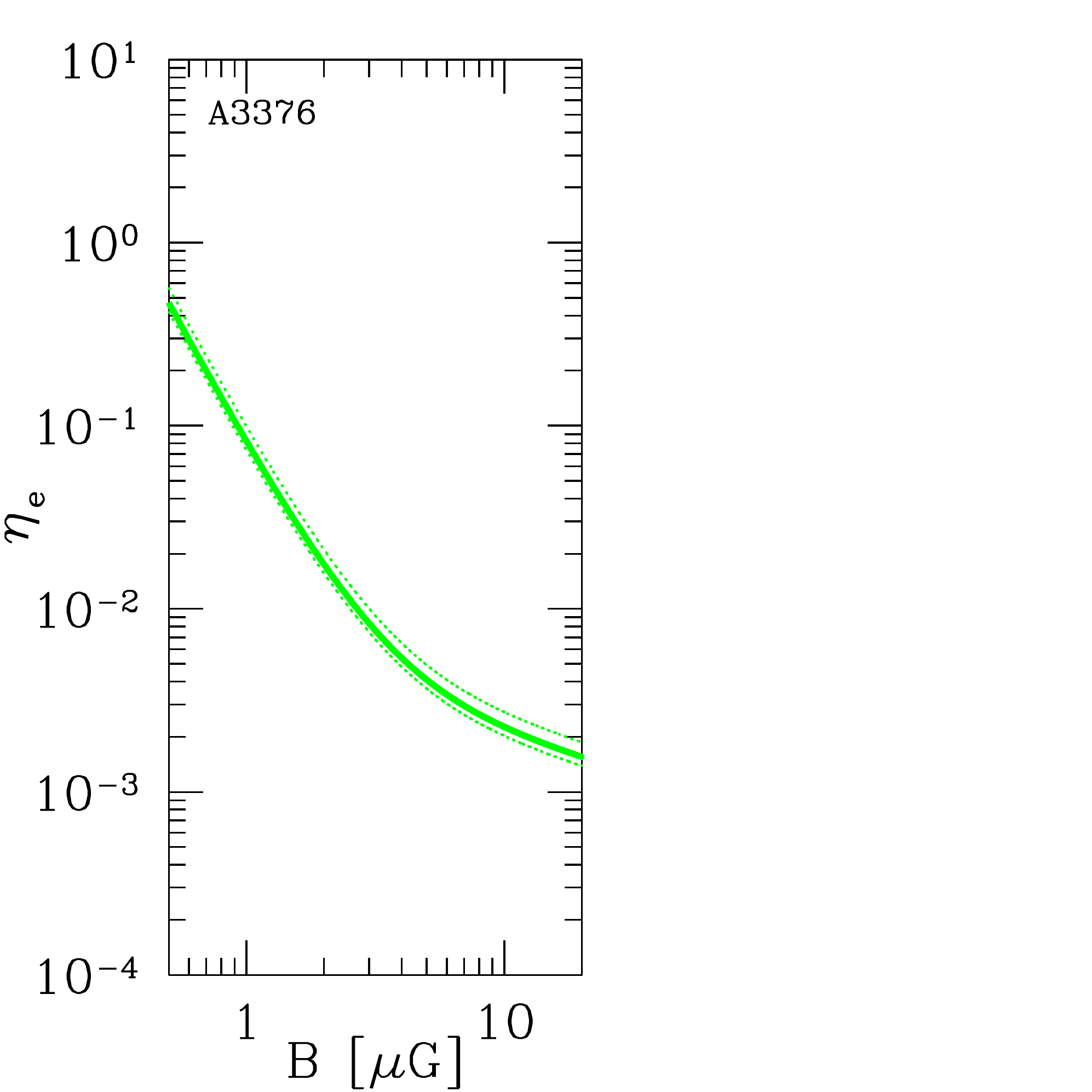} 
 \hspace*{-10mm} \\
 \vspace{5mm}
 \includegraphics[width=.24\textwidth,trim={0cm 0cm 5cm 0.8cm},clip]{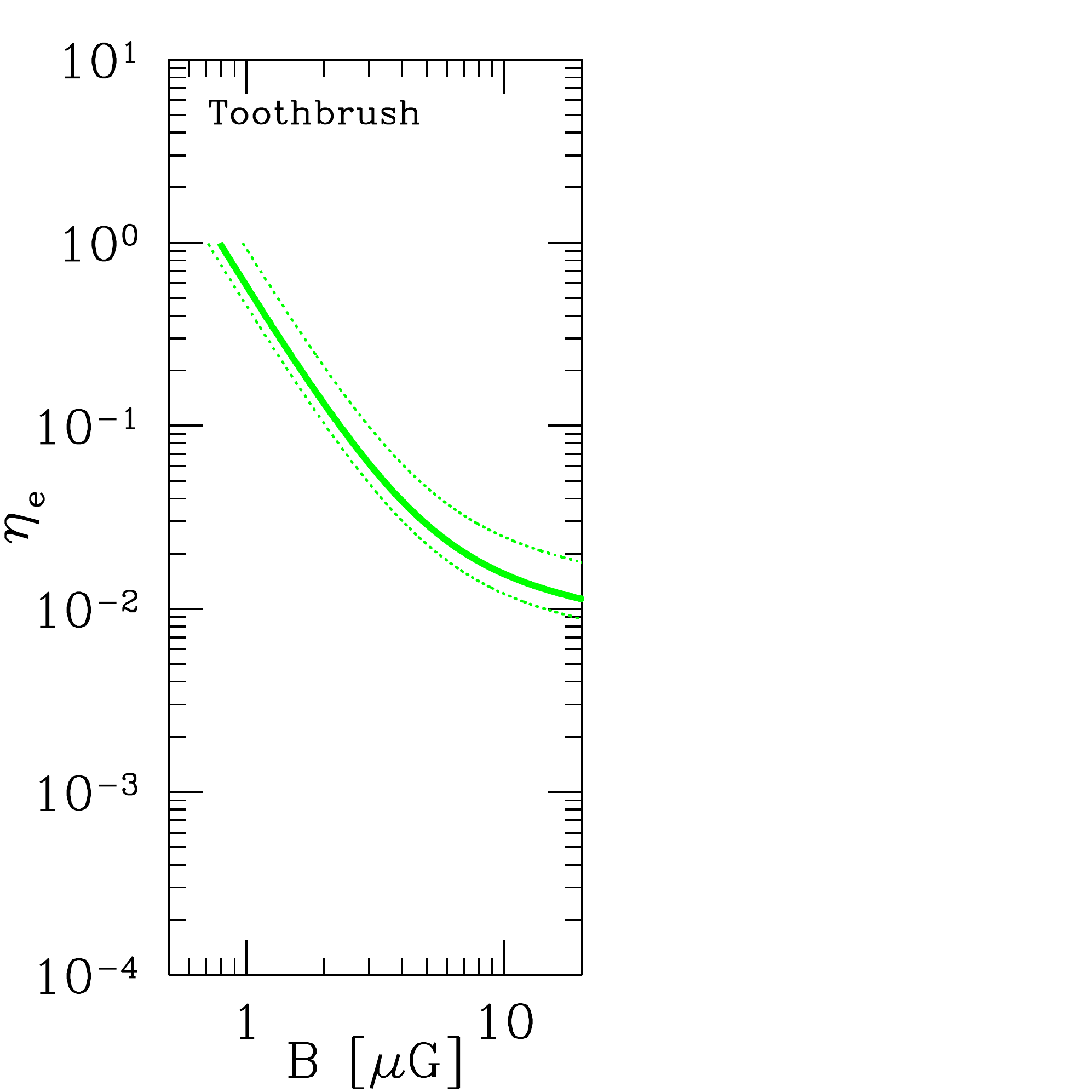}\hspace{-10mm}
 \includegraphics[width=.24\textwidth,trim={0cm 0cm 5cm 0.8cm},clip]{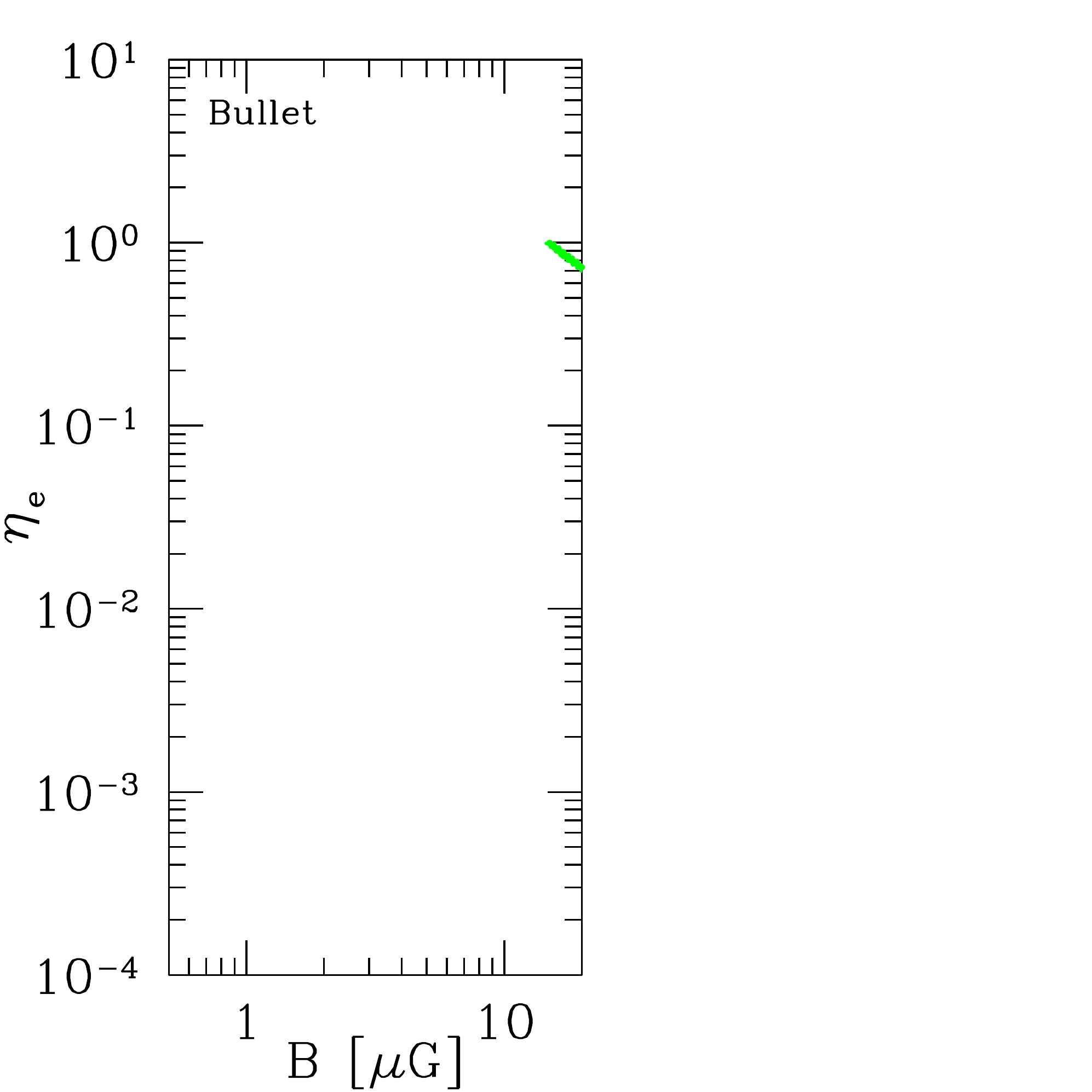}\hspace{-10mm} 
 \includegraphics[width=.24\textwidth,trim={0cm 0cm 5cm 0.8cm},clip]{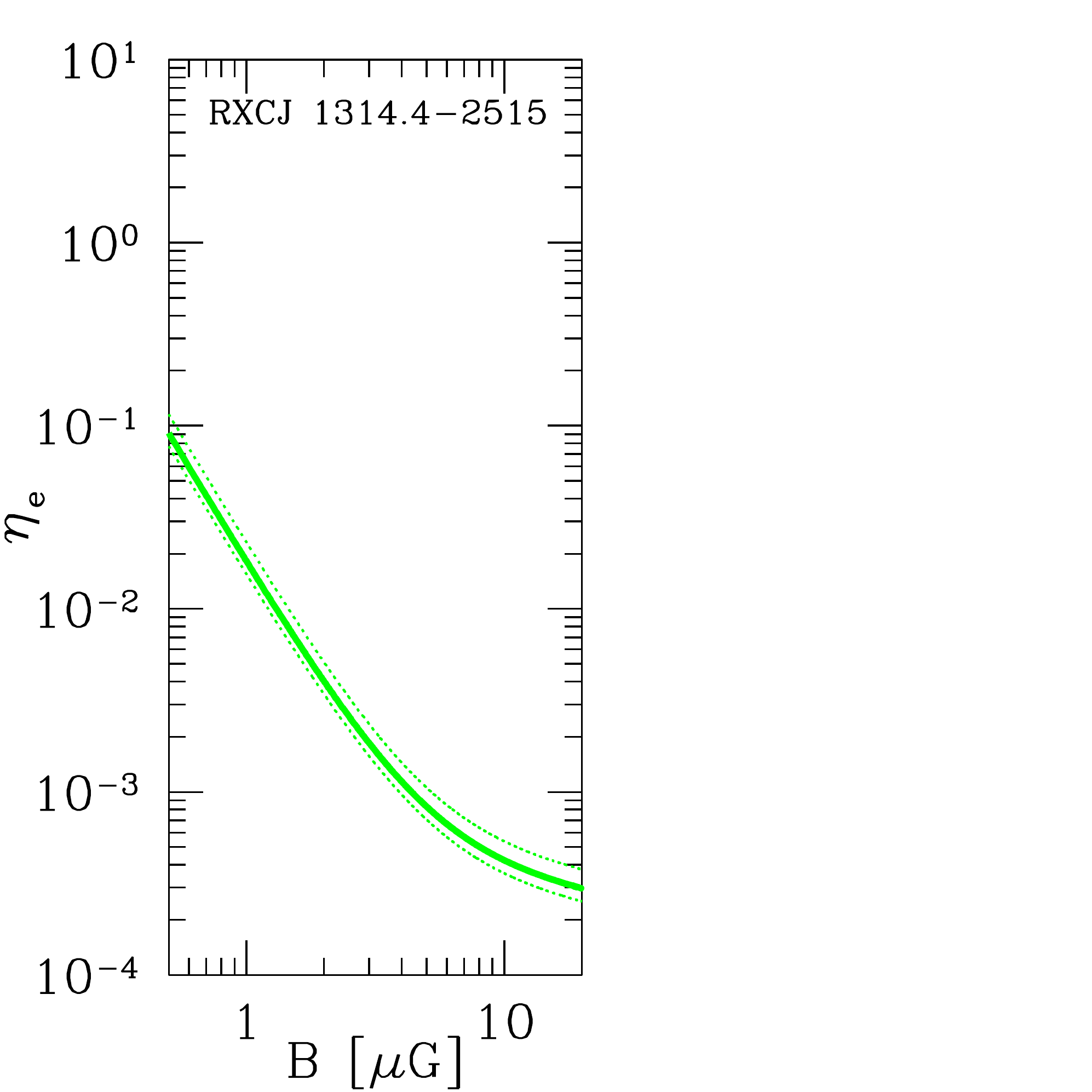}\hspace{-10mm}
 \includegraphics[width=.24\textwidth,trim={0cm 0cm 5cm 0.8cm},clip]{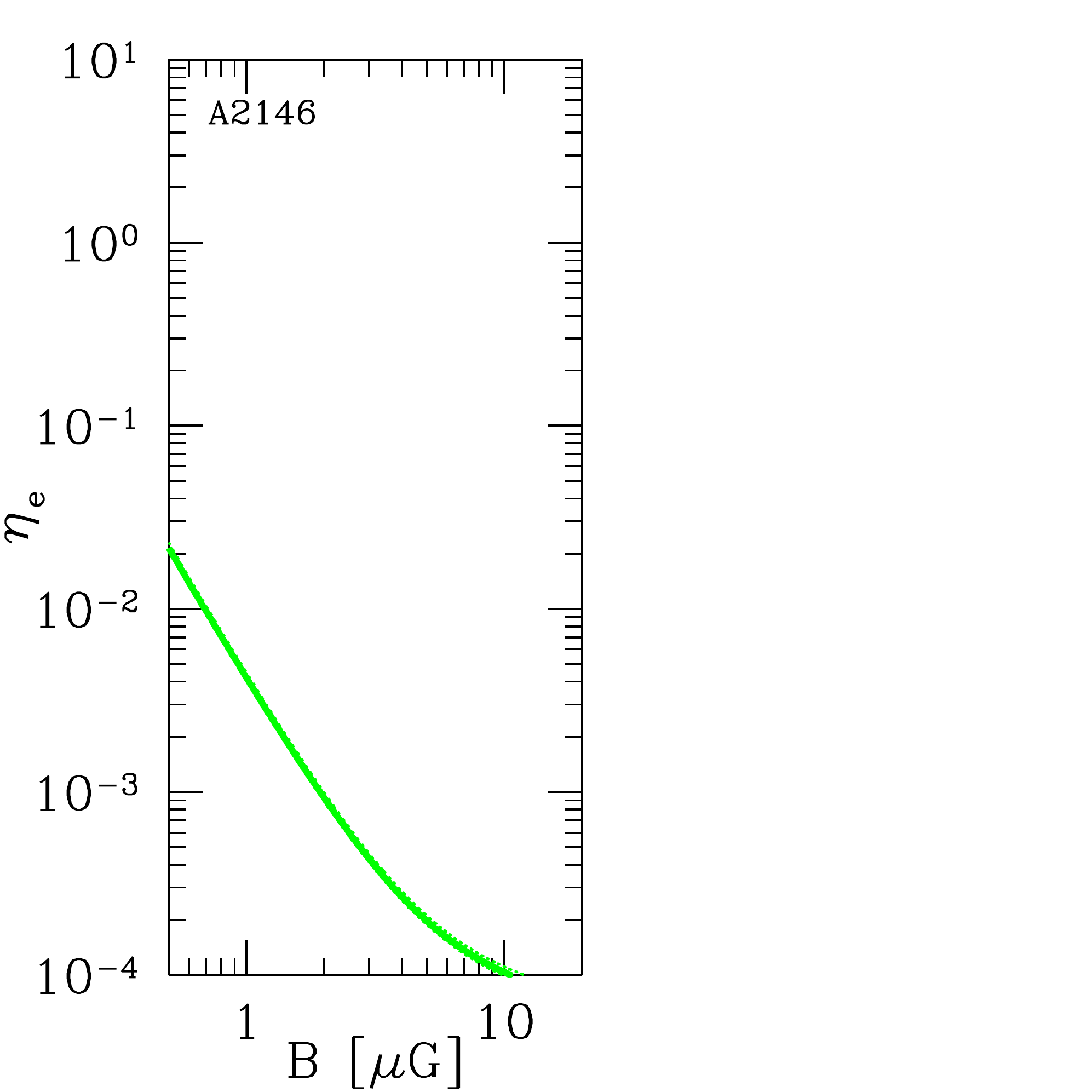}\hspace{-10mm}
 \includegraphics[width=.24\textwidth,trim={0cm 0cm 5cm 0.8cm},clip]{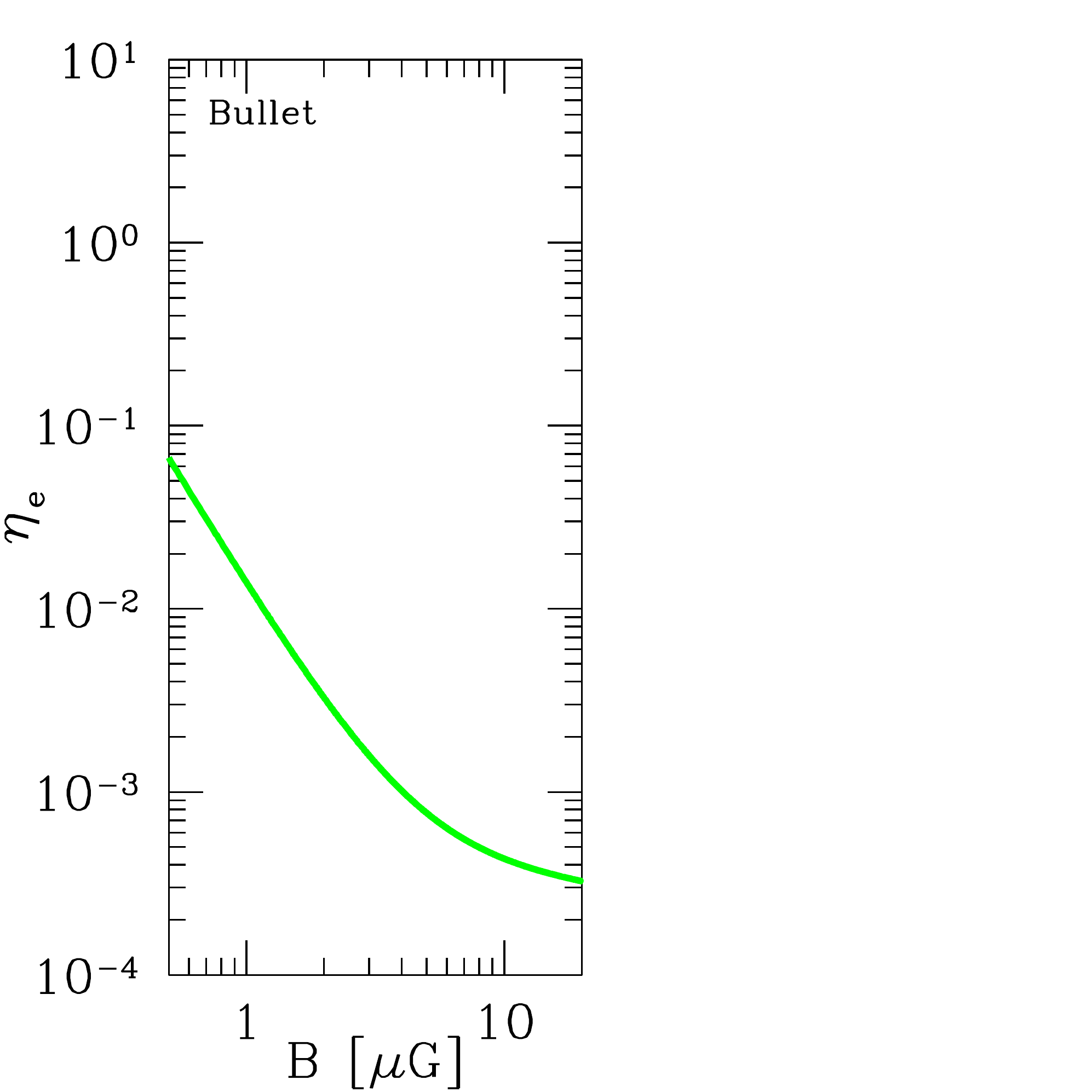}
 \hspace{-10mm}
 \caption{Same as Fig.~\ref{fig:eff_total} where now the {green} lines denote the case where the electron acceleration efficiency was computed adopting the Mach number measured in the X-rays (with its error bounds) and the central value of integrated spectral index measured in radio (\cf\ Table~\ref{tab:x_vs_radio}).}
 \label{fig:eff_hybrid}
\end{figure*}

Another \textit{ad hoc} possibility is shock acceleration from the thermal pool where the spectrum differs from the classical expectations based on the DSA theory. To explore this possibility, we evaluate the acceleration efficiencies assuming the values of Mach number and radio spectrum measured from X-ray and radio observations, respectively. Results are shown in Fig.~\ref{fig:eff_hybrid}; the acceleration efficiencies are similar to those computed assuming $\mach_{DSA}$ and $f_{DSA}=1$. We find that such a modified DSA model can explain the properties of the radio relics in our sample and their connection with the underlying shocks observed with $B = 5$ \muG\ and $\eta_e < 0.1$ except for the case of El Gordo, A521, and the Bullet Cluster; this latter is the most critical case due to the very steep synchrotron spectrum of the relic. The point here is to understand why clusters shocks, contrary to shocks in SNRs, produce a spectrum that is different from that implied by DSA. One possibility that is worth mentioning is that the transport of electrons across the shock is not diffusive but is superdiffusive, as recently proposed by \citep{zimbardo18} for example. However, at this stage this hypothesis is rather speculative.


\section{Conclusions}

We selected ten well-studied radio relics with detected underlying shocks to study the electron acceleration efficiency that is required to explain the origin of the diffuse radio emission. Our aim was to test whether or not DSA from the thermal pool is a general mechanism capable of explaining radio relics. Under this condition, the accelerated electrons are injected with a power-law momentum distribution from a minimum momentum $p_{min}$. Theoretically, this momentum is difficult to determine, and therefore we adopted the reverse approach to calculate $p_{min}$ and efficiency that would be necessary to explain the observed radio assuming DSA and shock parameters. \\
\indent
In general, the slope of the momentum distribution of the emitting electrons in our relics does not match DSA prediction if we assume the Mach number of the underlying shocks; this confirms previous findings \citep[e.g.,][]{akamatsu17a2255, urdampilleta18}. As a consequence we adopted two approaches to test DSA of thermal particles. Firstly, we assumed the Mach number measured in the X-rays: this model is based on the X-ray measurements but it only matches the radio spectrum of three relics in our sample. We find that this Mach number is generally too low -- typically $\mach_{X} \lesssim 2-2.5$ -- to reproduce the observed synchrotron luminosity of our radio relics, pinpointing a fundamental energy argument against this scenario. Under the optimistic assumption of a large $B$, only A521 and El Gordo can be explained by this mechanism assuming $\eta_e < 0.1$. We also show that the kinetic energy flux of the shocks constrained by the X-ray observations of our relics does not correlate with the observed radio luminosity, further disfavoring a direct connection between DSA of thermal electrons and radio relics. \\
Secondly, we assumed the value of the shock Mach number measured in X-rays but also assumed that the Mach number across the shock surface is not uniform and that the acceleration of electrons is dominated by the regions where \mach\ is higher. More specifically, we assumed that in these cases the acceleration is associated with Mach numbers that are implied by DSA from the spectrum of the radio relics. These Mach numbers for our relics are significantly larger than those measured in the X-rays and we used numerical simulations to correct for the fraction of shock surface that is occupied by these Mach numbers, $f_{DSA}$. This fraction is small, implying that the acceleration efficiency that is required to match the observed radio luminosity is still relatively large in these regions. Under the optimistic conditions of large $B$, only A3376 can be explained under the assumption of $\eta_e < 0.01$, whereas the other three relics (A2744, El Gordo, RXC J1314.4-2515) can be explained if $\eta_e < 0.1$. \\
Our main conclusion is therefore that, even if models of DSA can be reconciled with radio and X-ray measurements of Mach numbers, in general they fail to reproduce the observed synchrotron luminosity of radio relics due to the small amount of energy dissipated into particle acceleration at cluster shocks. \\
\indent
For this reason we also investigated a ``hybrid'' case where the acceleration efficiency is evaluated assuming the Mach number and synchroton spectrum derived from X-ray and radio observations. In this case the majority of radio relics in our sample can be explained in the case of high $B$ and assuming $\eta_e<0.1$, making this solution very appealing. Nonetheless, invoking a scenario that produces synchrotron spectra that differ from DSA predictions is rather speculative at present and requires further theoretical follow-up studies. \\
\indent 
The main conclusion of the present study is that other mechanisms, such as shock re-acceleration of supra-thermal seed electrons or a modification of standard DSA, are required to explain the formation of radio relics. The assumption of mildly relativistic electrons (\ie,\ higher $p_{min}$) would lower the required acceleration efficiencies and explain the luminosity of relics at weak shocks through re-acceleration mechanisms, as we discussed for A115 in an earlier work \citep{botteon16a115}.

\begin{acknowledgements}
We thank the anonymous referee for constructive comments which helped to improve the manuscript. We thank D. Dallacasa and F. Gastaldello for comments on an early version of the manuscript and S. Giacintucci, D. Hoang, R. Kale, T. Shimwell, T. Venturi and R. van Weeren for kindly providing us the displayed radio contours. GB acknowledges partial support from ASI/INAF n. 2017-14-H.O. The work of DR and SR was supported by the NRF of Korea through 2016R1A5A1013277 and 2017R1A2A1A05071429. The scientific results reported in this article are based in part on observations made by the \chandra\ X-ray Observatory and in part by \xmm, an ESA science mission with instruments and contributions directly funded by ESA Member States and NASA.
\end{acknowledgements}

%
%

\bibliographystyle{aa}
\bibliography{library.bib}

\begin{appendix}

\section{Details of the X-ray analysis}\label{app:param}

\begin{figure*}[ht]
 \centering
 \includegraphics[width=\textwidth,trim={0cm 0cm 0cm 0cm},clip]{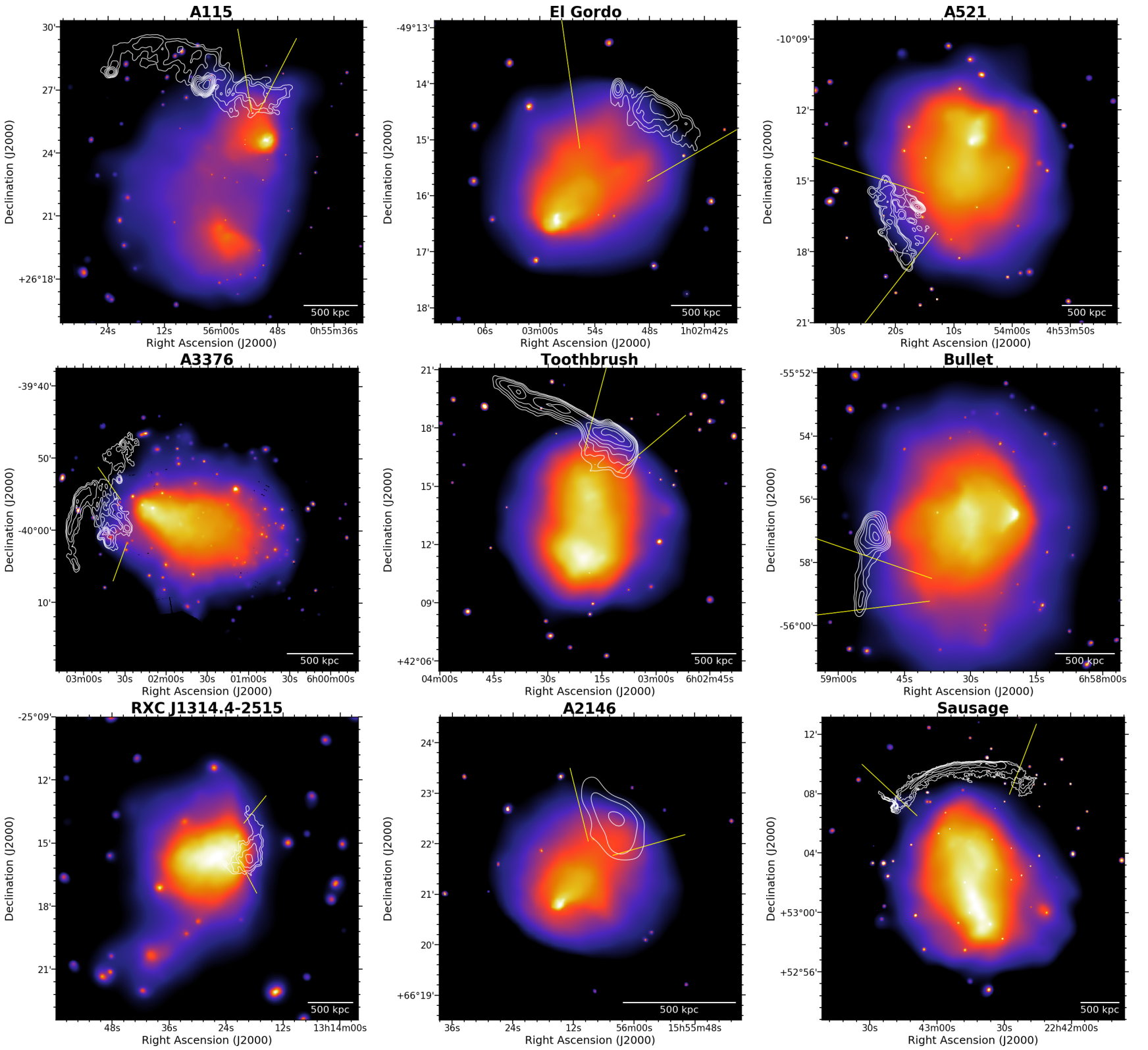} 
 \caption{X-ray images with overlaid radio contours from the works listed in Table~\ref{tab:sample_eff} and the sectors used for the analysis of the shock fronts.}
 \label{fig:cluster_view}
\end{figure*}

\begin{figure*}[ht]
 \centering
 \includegraphics[width=\textwidth,trim={0cm 0cm 0cm 0cm},clip]{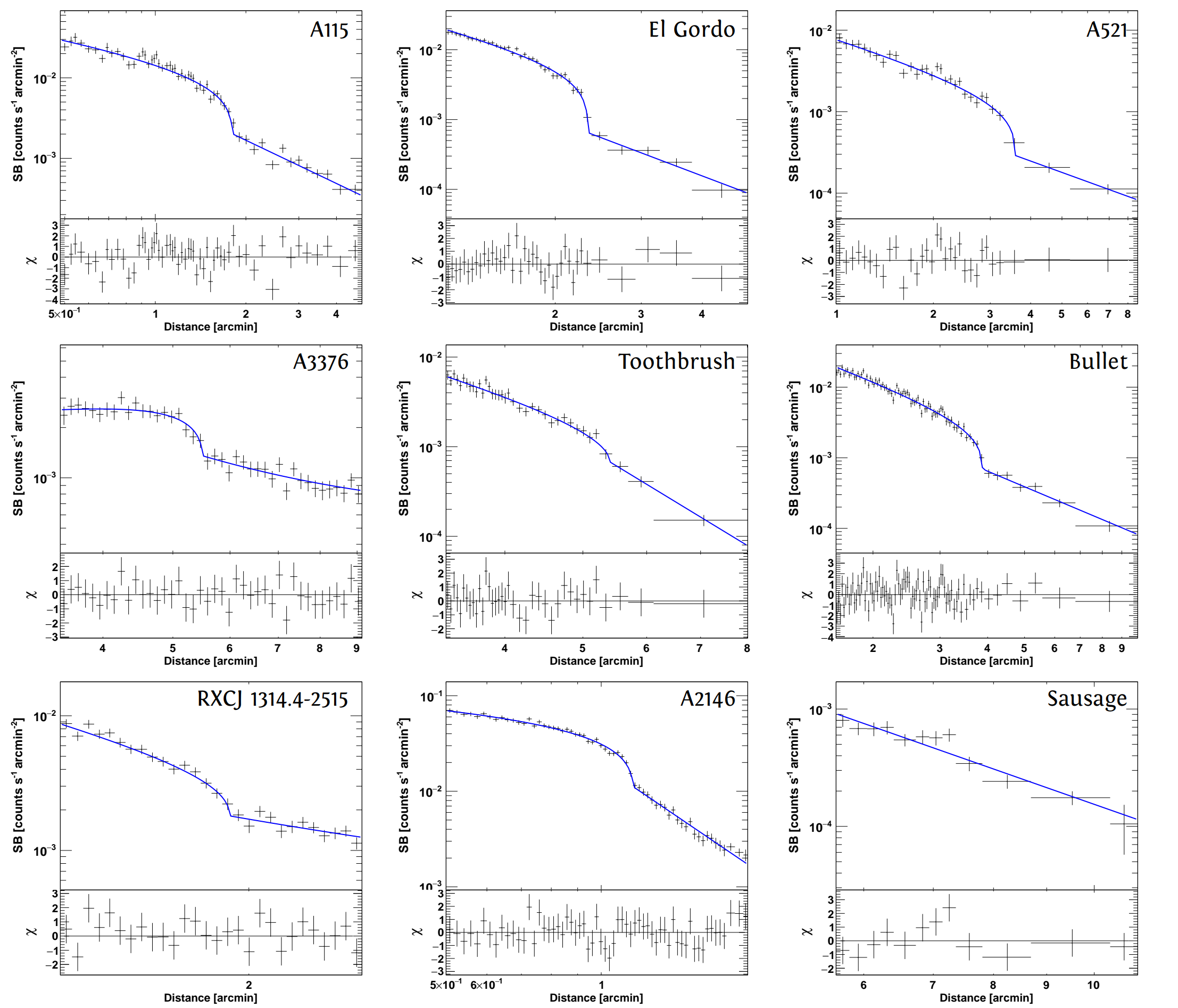} 
 \caption{X-ray surface brightness profiles extracted across the relics in the sample. A broken power-law model was used to fit the data for all the relics but the Sausage.}
 \label{fig:sb_total}
\end{figure*}

We retrieved all the \obsid s available for the clusters of our sample from the \chandra\footnote{\url{http://cda.harvard.edu/chaser/}} and \xmm\footnote{\url{http://nxsa.esac.esa.int/}} archives. In the cases where the clusters have been observed with both instruments, we used \chandra\ because its higher angular resolution allows us to better characterize the sharp edges of the shocks and to excise the point sources more accurately (\cf\ Table~\ref{tab:sample_eff}). \\
\indent
We performed a standard data reduction using \ciao\ v4.10, \chandra\ \caldb\ v4.7.8 and the \esas\ tools developed within the \xmm\ \sas\ v16.1.0. After the excision of contaminating point sources, surface brightness profiles across the shocks were extracted from the $0.5-2.0$ \kev\ exposure-corrected images of the clusters and fitted with \proffit\ v1.5 \citep{eckert11}. An underlying broken power law with a density jump was assumed to fit the data, which were rebinned to reach a minium S/N of 7. \proffit\ performs a modeling of the 3D density profile that is numerically projected along the line of sight under spherical assumption \citep[following the Appendix in][]{owers09sample}. Deprojected density profiles were recovered from the emission measure of the plasma evaluated in the case of an absorbed APEC model \citep{smith01} with metallicity assumed to be $0.3$ \zsun\ and total (\ie,\ atomic + molecular) hydrogen column density measured in the direction of the clusters fixed to the values of \citet{willingale13}. The choice of the soft band $0.5-2.0$ \kev\ ensures that the bremsstrahlung emissivity is almost independent of the gas temperature for $kT \gtrsim 3$ \kev\ \citep[e.g.,][]{ettori13rev}. \\
\indent
For the Sausage relic, no surface brightness discontinuity has been detected in the X-rays \citep{ogrean13sausage, ogrean14sausage}. Indeed, this is the only case where we used a single power-law model to fit the surface brightness profile. In this case, we used the density measured at the location of the relic from the single power-law model and assumed different Mach numbers in the analysis. \\
\indent
In Fig.~\ref{fig:cluster_view} we show the \chandra\ and \xmm\ images in the $0.5-2.0$ keV band of the clusters in the sample while in Fig.~\ref{fig:sb_total} we report the X-ray surface brightness profiles extracted across the relics considered in the analysis.

\end{appendix}

\end{document}